\documentclass[aps,prd,nopacs,floatfix,notitlepage,superscriptaddress,nofootinbib,twocolumn,a4paper,longbibliography]{revtex4-1}
\usepackage{amsfonts,amsmath,units,wasysym,epsfig,graphicx,verbatim,color,subfigure,graphicx,bm,mathrsfs,lipsum,hyperref}
\usepackage[utf8]{inputenc}
\usepackage[normalem]{ulem}  % \sout{old text} for strikeout

\newcommand{\caen}{\affiliation{Université de Caen Normandie, ENSICAEN, CNRS/IN2P3, LPC Caen UMR6534, F-14000 Caen, France}}
\newcommand{\iuf}{\affiliation{Institut Universitaire de France (IUF)}}
\newcommand{\stras}{\affiliation{Observatoire astronomique de Strasbourg, CNRS, Universit\'e de Strasbourg, 11 rue de l'Université, 67000 Strasbourg, France}}

\newcommand{\USAL}{\affiliation{Departamento de F\'isica Fundamental and IUFFyM, Universidad de Salamanca, Plaza de la Merced S/N, E-37008 Salamanca, Spain}}
\newcommand{\Uliege}{\affiliation{Space Sciences, Technologies and Astrophysics Research (STAR) Institute, Universit\'e de Li\`ege, B\^at. B5a, 4000 Li\`ege, Belgium}}

\newcommand{\brussels}{\affiliation{Institut d’Astronomie et d’Astrophysique, Universit\'e Libre de Bruxelles, CP 226, B-1050 Brussels, Belgium}}
\newcommand{\parisCit}{\affiliation{LUX, CNRS, Observatoire de Paris, Universit\'e PSL, Sorbonnes Universités, 5 place Jules Janssen, 92195 Meudon, France}}

\begin{document}

\title{Properties of neutron stars with hyperons within a generalized relativistic approach}
\author{Prasanta Char} \email{prasanta.char@usal.es } \USAL \Uliege
  \author{Chiranjib Mondal} \email{chiranjib.mondal@ulb.be}  \brussels
  \author{Timothé Alezraa}
\affiliation{Institut de Physique des deux Infinis de Lyon, Université Claude Bernard Lyon 1, 4 rue Enrico Fermi, 69622 Villeurbanne, France}
\author{Francesca Gulminelli} \email{gulminelli@lpccaen.in2p3.fr} \caen\iuf
    \author{Micaela Oertel} \email{micaela.oertel@astro.unistra.fr} \stras\parisCit

\begin{abstract}
In this work, we study the effects of $\Lambda$-hyperons on neutron star properties employing a  generalized relativistic framework for the equation of state (EOS). Different choices for defining the hyperonic couplings with different levels of parametric freedom are discussed. In all models, the predicted NS maximum masses are reduced compared with the purely nucleonic composition as expected. In the case of relating hyperonic couplings via $SU(6)$-symmetry arguments to the nucleonic ones, we find that NS radii for intermediate mass stars are shifted to higher values compared with purely nucleonic stars, in agreement with the existing literature. However, allowing for more freedom for the hyperonic couplings, the effect is strongly reduced, and the distributions in the NS mass-radius plane of models with and without hyperons become very close.
We have also investigated how different nucleonic density functionals influence the hyperon matter composition and neutron star properties.
\end{abstract}
\maketitle

\section{Introduction} 
Neutron stars (NSs) are remnants of violent core-collapse events of massive stars at the end of their life cycles \cite{Shapiro:1983du}. They have been observed in radio, X-rays, $\gamma$-rays, and  gravitational waves (GW) from binary neutron star (BNS) merger events \cite{Rezzolla:2018jee}. The density of the NS core may reach about a few times the nuclear saturation density ($n_{\mathrm{sat}}$). Hence, the composition of matter at that density is inaccessible to terrestrial experiments \cite{Glendenning:1997wn}. As the density increases inside the core of NS, the Fermi momenta and energies increase, too.  If we consider hyperons as ideal non-interacting fermions, following the Pauli exclusion principle, when the Fermi energy exceeds the rest mass energy of these heavier baryons, their appearances become energetically favorable \cite{ambartsumyan1960degenerate, Glendenning:1982nc, Glendenning:1984jr}. In this way, stable matter with strangeness degrees of freedom may exist inside NS. 

The emergence of these new particle species significantly influences the equation of state (EOS) and structure of NSs. Numerous studies have been conducted to understand and model such scenarios (see e.g. \cite{Chatterjee:2015pua, Tolos:2020aln, Burgio:2021vgk,Tong:2025sui}, and references therein 
%for detailed reviews
). As a general feature, the onset of hyperons at $\sim$ 2-3$n_{\mathrm{sat}}$ makes the EOS softer, which, in turn, makes the maximum attainable mass by the EOS smaller than that in absence of hyperons. However, the radio observations of massive pulsars provide evidence against extreme softening thereby ruling out several EOS models. Over the years, we have observed PSR J1614-2230 with $M = 1.908 \pm 0.016 M_\odot$ \cite{Demorest:2010bx, Fonseca:2016tux, Arzoumanian:2017puf}, PSR J0348–0432 with  $M = 2.01 \pm 0.04 M_\odot$ \cite{Antoniadis:2013pzd}, PSR
J0740+6620 with $M = 2.08 \pm 0.07 M_\odot$ \cite{Cromartie:2019kug, Fonseca:2021wxt}, providing us a strong benchmark for the NS maximum mass to qualify for a viable EOS. This has led to the so-called ``hyperon puzzle", which emphasizes the difficulties to reconcile the pulsar mass measurements with the incorporation of hyperons in NS matter. 
 This is particularly true for ab-initio calculations of the EOS with hyperons. However, these calculations rely on the knowledge of scattering phase shifts in the vacuum in a large energy domain. Such data are scarce for nucleon-hyperon scattering, and not available in the hyperon-hyperon case \cite{Burgio:2021vgk}. This calls for more phenomenological approaches where the behavior of nucleons and hyperons in dense matter is directly governed by  phenomenological coupling constants that should be fixed on experimental or observational data. In general, for the latter approaches based on (non-)relativistic energy density functionals are employed. In this context,
 one can use Bayesian techniques to optimize the parameter space of the EOS functional incorporating hyperons using different  constraints from (hyper-)nuclear physics and astrophysics. Among others, the GW data from BNS merger event GW170817 reported by the LVK collaboration provides the tidal deformability  that can constrain NS radii~\cite{TheLIGOScientific:2017qsa,LIGOScientific:2018cki,LIGOScientific:2018hze}. Systematic comparisons between nucleonic and hyperonic stars within different types of relativistic mean-field model  suggest that somewhat stiffer nucleonic EOSs are favored to sustain the softening after the appearance of hyperons resulting in a general increase of the radii of the stars in the M-R sequence \cite{Traversi:2020aaa,Ghosh:2022lam,Malik:2022jqc}. Simultaneous mass and radius measurements of pulsars PSR J0030+0451 \cite{Riley:2019yda, Miller:2019cac}, PSR J0740+6620 \cite{Riley:2021pdl, Miller:2021qha}, and PSR J0437-4715 \cite{Choudhury:2024xbk} using X-ray data by NICER collaboration may also provide further constraints on the M-R space. Combined effects of GW and NICER constraints have been investigated on hyperon models for NSs  
in recent works \cite{Sun:2022yor, Huang:2024rvj}. 
Notably, the study of ref.\cite{Huang:2024rvj} underlines the importance of the PSR J0437-4715 \cite{Choudhury:2024xbk} measurement which favors lower radii for a canonical 1.4 $M_o$ NS. Still, the existing constraints are not sufficient to draw quantitative conclusions on the importance of the hyperon component in the NS core. \\
 In view of future observations, possibly with third generation GW interferometers \cite{2025arXiv250312263A,Branchesi:2023mws,ET:2019dnz,Evans:2021gyd}, it is important to explore the general features and possible behaviors of the EOS with the inclusion of hyperon degrees of freedom. 
Because of the huge uncertainties on the high density hadronic EOS, especially when introducing strangeness, the existing works in the literature cannot yet propose a complete exploration of the parameter space. In particular, only a limited set of nucleonic models is often considered \cite{Huang:2024rvj}, or strong hypothesis are made on the behavior of the hyperon couplings \cite{Malik:2022jqc}. \\
In this work, we perform a full Bayesian study of the hadronic EOS exploring its variability both in the nucleonic and hyperonic sector, as allowed by present experimental and observational constraints. 
We consider density-dependent couplings in the Relativistic Mean Field (RMF) approach, and compare the results obtained using two different functional forms for the density-depen%A recent work on calculating the hyperon single-particle potential for infinite nuclear matter from chiral hyperon-nucleon interactions suggests a larger range for ${\cal U}_\Lambda^N(n_{sat})$, with values approximately between $-25$ and $-60$ MeV \cite{Jinno:2025vgm}, thus in particular the possibility of a stronger attraction than assumed here. Studying the implications of these findings in the context of neutron star matter will be kept for future work.
dence. Concerning the hyperon couplings, we allow for both attractive and repulsive hyperon-hyperon interactions, and study in detail the effect of relaxing the simple SU(6) hypothesis, thus effectively decoupling the behavior of strange and non-strange baryons in dense matter. To keep a treatable dimension of the parameter space, we consider only the lightest $\Lambda$ hyperons, where the $\Lambda$-nucleon interaction is the one for which the most relevant constraints can be applied from hypernuclear experiments. Detailed compilations of the present phenomenological constraints on the various hyperon couplings, illustrating the large uncertainties still present for $\Sigma$ and $\Xi$ interactions, can be found in recent literature, for instance in \textcite{Fortin:2020qin}. Nevertheless, our predictions concerning the $\Lambda$ fraction should be taken with care, and rather considered as a qualitative prediction of the global strangeness content of the star.

We adopt a generalized approach in the sense of Refs. \cite{Char:2023fue, Scurto:2024ekq}. Rather than calibrating a single interaction, we use the relativistic density functional as a generator of a large ensemble of models spanning the empirical nuclear matter parameter (NMP) space. The structure of the paper is organized as follows. In Section \ref{sec:formalism}, we discuss our approach, the underlying density functional in \ref{sec:Lagrangian}, the choices of coupling parameters of nucleons and hyperons in \ref{sec:coupling_par}, and then explain the concept of stability of the EOSs at the onset of hyperons in \ref{sec:stability}. Next, we discuss our Bayesian methodology in Section \ref{sec:bayes} with the outline of our constraints. We present our results in Section \ref{sec:results}, with  a discussion of the effect of relaxing the $SU(6)$ hypothesis for the hyperonic couplings in \ref{sec:compare_hyp}, a comparison between nucleonic and hyperonic models in \ref{sec:compare_nuchyp}, and a comparison of hyperonic matter with different underlying nucleonic functionals in \ref{sec:compare_su6}. Finally, we summarize our conclusions in Section \ref{sec:conclusion}.

\section{Formalism}
\label{sec:formalism}

\subsection{Effective Lagrangian}
\label{sec:Lagrangian}
Because of the strong nature of the couplings between quarks and gluons, starting from quantum chromodynamics, the description of dense matter is not straightforward in the non-perturbative regime. Even in the modern era of supercomputing facilities, realistic calculations are extremely demanding and not possible at NS densities. With baryonic degrees of freedom, relativistic density functionals have provided a feasible alternative to describe matter across a wide range of densities relevant for both finite nuclei and NSs.

In the present work, we have used the baryonic model Lagrangian density, ${\cal L}_B$, of the form
\begin{widetext}
\begin{eqnarray}
\label{eq_lag_b}
{\cal L}_B &=& \sum_{B= N, \Lambda} \bar\psi_{B}\left(i\gamma_\mu{\partial^\mu} - m_B
+ \Gamma_{\sigma B} \sigma - \Gamma_{\omega B} \gamma_\mu \omega^\mu 
-  \Gamma_{\rho B} 
\gamma_\mu{\mbox{\boldmath $\tau$}}_B \cdot 
{\mbox{\boldmath $\rho$}}^\mu  \right)\psi_B\nonumber\\
&+& \frac{1}{2}\left( \partial_\mu \sigma\partial^\mu \sigma
- m_\sigma^2 \sigma^2\right)
-\frac{1}{4} \omega_{\mu\nu}\omega^{\mu\nu} + \frac{1}{2}m_\omega^2 \omega_\mu \omega^\mu
- \frac{1}{4}{\mbox {\boldmath $\rho$}}_{\mu\nu} \cdot
{\mbox {\boldmath $\rho$}}^{\mu\nu}
+ \frac{1}{2}m_\rho^2 {\mbox {\boldmath $\rho$}}_\mu \cdot
{\mbox {\boldmath $\rho$}}^\mu.
\end{eqnarray} 
\end{widetext}
The baryons represented by the spinors $\psi_{B}$ include the nucleons and $\Lambda$-hyperons, which interact through isoscalar-scalar $\sigma$, isoscalar-vector $\omega$ and isovector-vector $\rho$ mesons.  ${\mbox{\boldmath $\tau_{B}$}}$ is the isospin operator. The field strength tensors for the vector mesons are given by
$ \omega^{\mu \nu}= \partial^ \mu \omega^ \nu-\partial^\nu \omega^ \mu$
and
${\boldsymbol \rho^{\mu \nu}}= \partial^ \mu {\boldsymbol \rho}^ \nu-\partial^\nu {\boldsymbol \rho}^ \mu $. An additional vector meson $\phi$ and a scalar meson $\sigma^*$ are also included 
which are only coupled to strange baryons
\cite{Schaffner:1993qj,Schaffner:1995th, Weissenborn:2011kb,Weissenborn:2011ut}\footnote{Please note that relaxing the $SU(6)$ hypothesis, the $\phi$ can in principle couple to nucleons, too, see e.g.~\cite{Weissenborn:2011ut,Oertel:2014qza}. For simplicity we will, however, assume $\Gamma_{\phi N} = 0$ throughout the present study.}. These interaction among $\Lambda$-hyperons can be represented  
by the Lagrangian density ${\cal L}_{\Lambda\Lambda}$ as

\begin{eqnarray} \label{eq_lag_y}
{\cal L}_{\Lambda\Lambda}&=& \bar\psi_{\Lambda}\left(\Gamma_{\sigma^* \Lambda} \sigma^* 
 - \Gamma_{\phi \Lambda} \gamma_\mu \phi^\mu
	\right)\psi_{\Lambda} \nonumber\\
&& + \frac{1}{2}\left( \partial_\mu \sigma^* \partial^\mu \sigma^*
- m_{\sigma^*}^2 {\sigma^*}^2\right) \nonumber\\
&& -\frac{1}{4} \phi_{\mu\nu}\phi^{\mu\nu}
+\frac{1}{2}m_\phi^2 \phi_\mu \phi^\mu~.
\end{eqnarray}

The $\Gamma$-s appearing in Eqs.(\ref{eq_lag_b},\ref{eq_lag_y}) are density-dependent coupling parameters in different channels. A more detailed discussion  is provided in the next subsection. Leptons are treated as non-interacting particles and described by the standard Lagrangian density ${\cal L}_l$ as
\begin{eqnarray}
{\cal L}_l &=& 
\sum_l \bar\psi_l\left(i\gamma_\mu {\partial^\mu} - m_l \right)\psi_l ~.
\end{eqnarray}
Here, $\psi_l$ ($l \equiv {e,\mu}$) is the lepton spinor.

\subsection{Coupling parameters}\label{sec:coupling_par}
In the present work, we have primarily used the model of GDFM type \cite{Gogelein:2007qa}, to parametrize the density-dependent coupling parameters. The corresponding functional form is given by     
\begin{equation}
 \Gamma_{iB}(n_B)=a_{iB}+(b_{iB}+d_{iB}\,x^3)e^{-c_{iB}\,x},
\label{eq:GDFM}
\end{equation}
 with B=N,$\Lambda$,  $i=\sigma, \omega, \rho, \sigma^*, \phi;\ x=n_B/n_0$, $n_0$ being a 
constant scaling density to the number density $n_B$. The parameter $n_0$ is different 
from the nuclear saturation density $n_{sat}$, but generally chosen close to. The ranges of the free parameters 
$a,b,c$ and $d$'s used for $\Gamma_{(\sigma,\omega,\rho)N}$ can be found in Table 
I of Ref. \cite{Char:2025zdy}. Concerning the hyperons,
\footnote{One should note here that for the hyperons, we 
are only concerned about the isoscalar channel, since we only consider the 
$\Lambda$-hyperons.}
we have used two distinctive ways to generate the interaction parameters:
\begin{itemize}
	     
\item In the first setting, we follow the
         simplest representation of hadrons via the valence quark model preserving both flavor and spin symmetry, where the meson-hyperon vector couplings are uniquely defined from the corresponding meson-nucleon couplings using the $SU(6)$ symmetry relations as is common practice in the literature~\cite{Weissenborn:2011ut,Malik:2022jqc,Fu:2022eeb}:       
        
\begin{eqnarray}
\label{eq:su6_1}
\frac{1}{2}\Gamma_{\omega \Lambda} = 
\frac{1}{3} \Gamma_{\omega N},\\
\label{eq:su6_2}
2 \Gamma_{\phi \Lambda} = 
-\frac{2\sqrt{2}}{3} \Gamma_{\omega N}.
\end{eqnarray}

	\item For the second, termed as ``Ratio", we fix the hyperonic parameters such that the associated couplings 
$\Gamma_{i\Lambda}$'s (with $i=\sigma, \omega, \sigma^*, \phi$) are proportional to  the nucleonic couplings. The corresponding ratios :
\begin{eqnarray}
    R_{\omega\Lambda} &=& \Gamma_{\omega \Lambda}/ \Gamma_{\omega N},\\
   R_{\phi\Lambda} &=& \Gamma_{\phi \Lambda}/ \Gamma_{\omega N}.\label{eq:ratio3}
\end{eqnarray}
 are randomly varied within the intervals reported in Table \ref{tab:parameters}.
 
\end{itemize}
Please note that the "Ratio" setting does not correspond to a fully free variation of the hyperonic couplings, since the density-dependence of the coupling parameter is fixed to that of the nucleonic ones, a full exploration of the parameter space being beyond the scope of the present paper.

 In principle, the scalar meson ($\sigma$) coupling to $\Lambda$-hyperons can be derived from the $A\to\infty$ extrapolation of the experimental binding
energy of single-$\Lambda$ hypernuclei. This gives a constraint on the $\Lambda$ potential $U_\Lambda^N$ in symmetric nuclear matter around saturation 
$U_\Lambda^N(n_{\mathit{sat}})\approx -30$ MeV \cite{HASHIMOTO2006564,Gal:2016boi}, where  $U_\Lambda^N$ is given by: 

\begin{equation}\label{eq:ulambda}
	{\cal U}_{\Lambda}^N(n_B) 
    = - \Gamma_{\sigma \Lambda} {\sigma} + \Gamma_{\omega \Lambda} {\omega_0} +\Sigma^{(r)}, 
\end{equation}
with the rearrangement term $\Sigma^{(r)}$ and $\sigma,\omega_0$ corresponding to the mean field expectation values for the meson fields at a given density. In symmetric nuclear matter the mean-field values and the re-arrangement term only receive nucleonic contributions.

However, because of the uncertainties in the hypernuclear data and the ambiguities in the definition of the relevant density, we have chosen to vary $\Gamma_{\sigma \Lambda}$ using a random variation of the ratio $R_{\sigma \Lambda} \equiv \Gamma_{\sigma \Lambda}/ \Gamma_{\sigma N}$ as for the vector couplings in the "Ratio" setting, and impose the experimental constraint on the $\Lambda$-potential at the level of the posterior (see Section \ref{sec:bayes}).  In both settings, for the $\sigma^*$-coupling similarly a random sampling of the ratio $R_{\sigma*\Lambda} = \Gamma_{\sigma* \Lambda}/ \Gamma_{\sigma N}$ is assumed, see also the discussion in Section~\ref{sec:stability}.

Finally, in order to compare with the existing literature, we have also extended our analysis with 
another type of density-dependence, designated as TW \cite{Typel:1999yq}. This amounts to replace Eq.(\ref{eq:GDFM}) by :
\begin{eqnarray}
\Gamma_{iN}(n_B) &=& \Gamma_{iN}(n_{\mathrm{sat}})f_{iN}(y),\quad\mbox{with}\quad \\
f_{iN}(y) &=& a_{iN}\frac{1+b_{iN}(y+d_{iN})^2}{1+c_{iN}(y+d_{iN})^2},
\label{eq:gamadefault}
\end{eqnarray}
for $i=\sigma, \omega;$, and
\begin{eqnarray}\label{eq:TWrho}
\Gamma_{\rho N}(n_B)=\Gamma_{\rho N}(n_{\mathrm{sat}})e^{-a_{\rho N}(y-1)},
\end{eqnarray}
where $y=n_B/n_{\mathrm{sat}}$.
The specific calculations using the TW  couplings will be aimed at setting the possible model dependence of the results due to the choice of the functional form for the density-dependence. For this reason, concerning the hyperon couplings, we will  restrict ourselves to the $SU(6)$ case, for which the nucleonic coupling parameters totally enclose the vector hyperonic interactions through  Eqs. (\ref{eq:su6_1}),(\ref{eq:su6_2}). 

At this point, we clarify our use of the term ``generalized relativistic modeling.'' It is inspired by the nucleonic metamodel of Refs. \cite{Margueron:2017eqc, Margueron:2017lup, Margueron:2018eob}, in which a functional, parametrized by the empirical NMPs, is used not to fix a single interaction, but to generate a broad ensemble of models spanning the allowed NMP space. It is thus a physics-informed generative framework, not an agnostic mathematical expansion of the EOS. Following Refs. \cite{Char:2023fue, Scurto:2024ekq} (see also \cite{Char:2025zdy}), we extend this idea to relativistic density functionals, but for the moment without checking explicitly that our approach can faithfully reproduce the majority of existing models as done for the metamodel~\cite{Margueron:2017eqc,Margueron:2017lup,Margueron:2018eob}. The density-dependent functional, with all the couplings left free, serves as a parent structure that we use to span the NMP space, retaining only those parameter sets whose NMP combinations satisfy the bound dictated by nuclear and crust physics. This differs both from a standard RMF model (e.g., DD2, DDME2), which denotes a single fixed set of couplings, and from a Bayesian optimization of one such model. Here, the functional generates a large ensemble of distinct valid models rather than a best fit. It is this generative, NMP spanning character that we designate by the term  "generalized" description.

\begin{table}
    \centering
    \begin{tabular}{ccc}
    \hline
    \hline
          Parameters & Minimum value & Maximum value \\
        \hline 
        \hline
      & Ratio & \\
        \hline
        \hline
       $R_{\omega \Lambda}$ & 0.55 & 0.8 \\ 
        $R_{\phi \Lambda}$ & -0.7 & -0.2 \\
        $R_{\sigma\Lambda}$ & 0.55 & 0.7 \\
          $R_{\sigma* \Lambda}$ & 0.1 & 0.7 \\
        \hline
        \hline 
      & SU(6) & \\
        \hline
        \hline
        $R_{\sigma\Lambda}$ & 0.55 & 0.7 \\
        $R_{\sigma* \Lambda}$ & 0.1 & 0.7 \\
        \hline 
        \hline
    \end{tabular}
    \caption{Ranges of model parameters used to explore the distribution of SU(6) and Ratio cases for Bayesian studies. The parameter ranges for the nucleonic couplings are the same as in Refs.~\cite{Char:2025zdy,Char:2023fue}.}
     \label{tab:parameters}
\end{table}

\subsection{Stability at hyperon onset}\label{sec:stability}
As previously discussed in Refs.~\cite{Schaffner-Bielich:2000igu,Gulminelli:2012iq,Gulminelli:2013qr,Oertel:2014qza,Oertel:2016xsn}, the generic presence of attractive and repulsive couplings suggests the possible existence of a phase transition involving strangeness. In these works, it was shown that under the strangeness equilibrium condition of stellar matter, the onset of hyperons can appear via a ﬁrst-or second- order phase transition, depending on the detailed values of the coupling parameters within a chosen model.  The existence of a ﬁrst order phase transition can be spotted by analyzing the curvature of the thermodynamic potential
as a function of its extensive variables, indicating the presence of a
spinodal instability related to the phase transition.
Within a non-relativistic setup~\cite{Gulminelli:2012iq,Gulminelli:2013qr}, the parameter space for such an instability appears to be relatively large, whereas for relativistic density functionals it seems that such an instability requires a very strongly attractive hyperon-hyperon %$YY$ 
interaction at low densities -- here represented by the $\sigma^*$-$\Lambda$ coupling, incompatible with current experimental constraints~\cite{Oertel:2014qza}. However, only a few selected models have been considered in these works and we thus want to revisit the question here within the  generalized approach allowing for a systematic exploration of the parameter space.

Technically, in order to identify the unstable region, we perform a convexity analysis~\cite{Gulminelli:2012iq,Gulminelli:2013qr,Oertel:2014qza,Oertel:2016xsn,Avancini:2006zi,Ducoin:2005aa} of the total energy density $\varepsilon(\{n_i\})$,  which is the adequate thermodynamic potential at zero temperature. Here, $n_i$'s correspond to the number densities associated with good quantum numbers for the strong interaction which is responsible of the phase transition. In neutron star matter with electrical charge neutrality imposed, charge
is not a good degree of freedom~\cite{Providencia:2006mm,Chomaz:2005xn} and the relevant 
number densities are baryon number $n_B$, strangeness $n_S$ and (electronic) lepton number $n_L$~\cite{Gulminelli:2013qr}.
The system is then thermodynamically stable as long as all eigenvalues
of the curvature matrix, $C_{ij} = \frac{\partial^2 \varepsilon }{\partial n_i n_j}$, remain positive.  
We have thereby ${i,j}\in {B,S,L}$. For neutron star matter, we can restrict the analysis to a line in the three-dimensional density space with $\mu_S = 0$ and $\mu_L = 0$ corresponding to strangeness changing weak equilibrium and neutrinoless $\beta$-equilibrium. Let us stress that we do not assume weak equilibrium to be maintained throughout the fluctuations such that the analysis of the curvature matrix remains three-dimensional, see also the discussion in Ref.~\cite{Oertel:2014qza}. 

\section{Bayesian analysis}\label{sec:bayes}
\begin{figure*}
    \centering
    \begin{tabular}{cc}
    \includegraphics[width=0.48\textwidth]{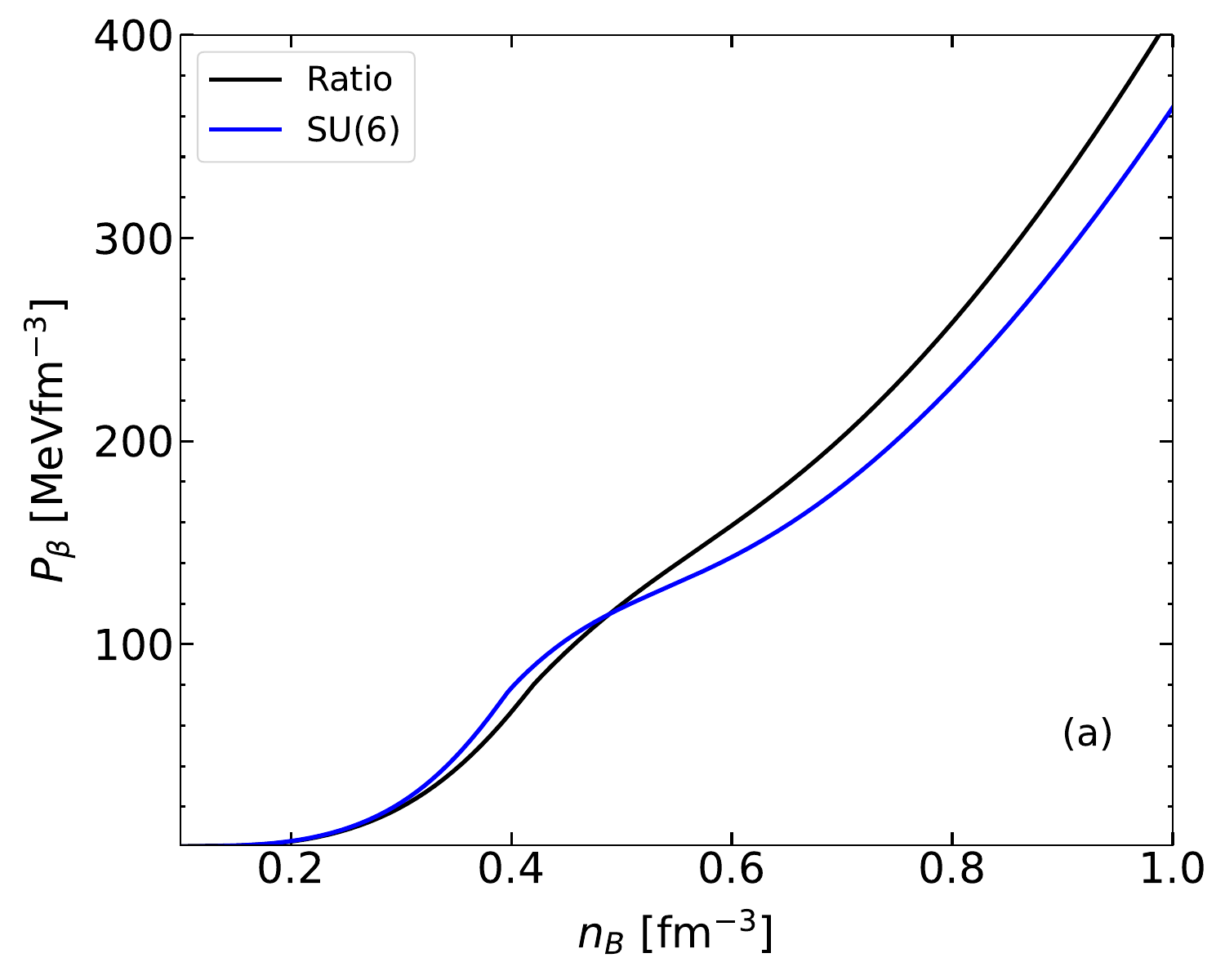} &  \includegraphics[width=0.48\textwidth]{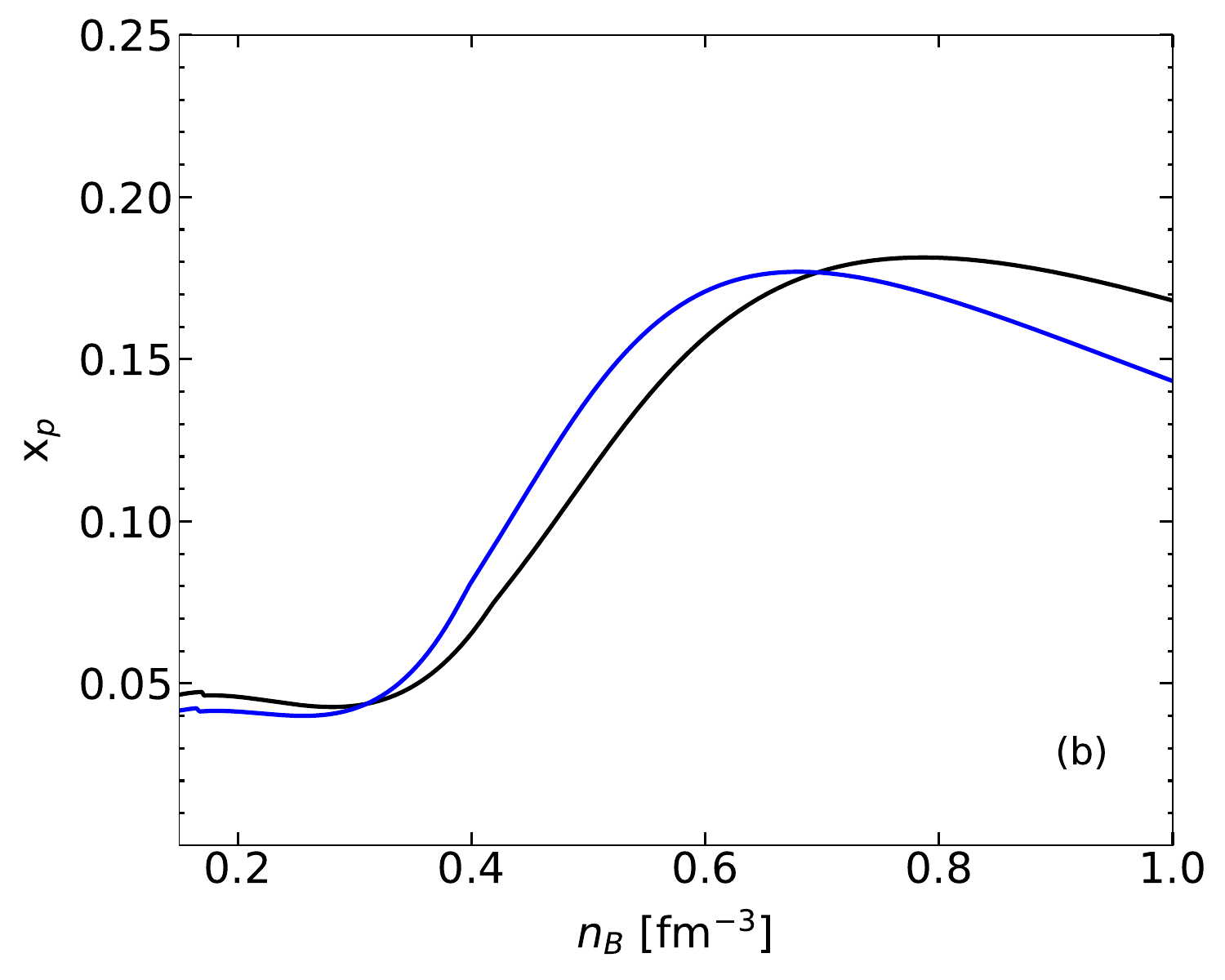} \\
	\includegraphics[width=0.48\textwidth]{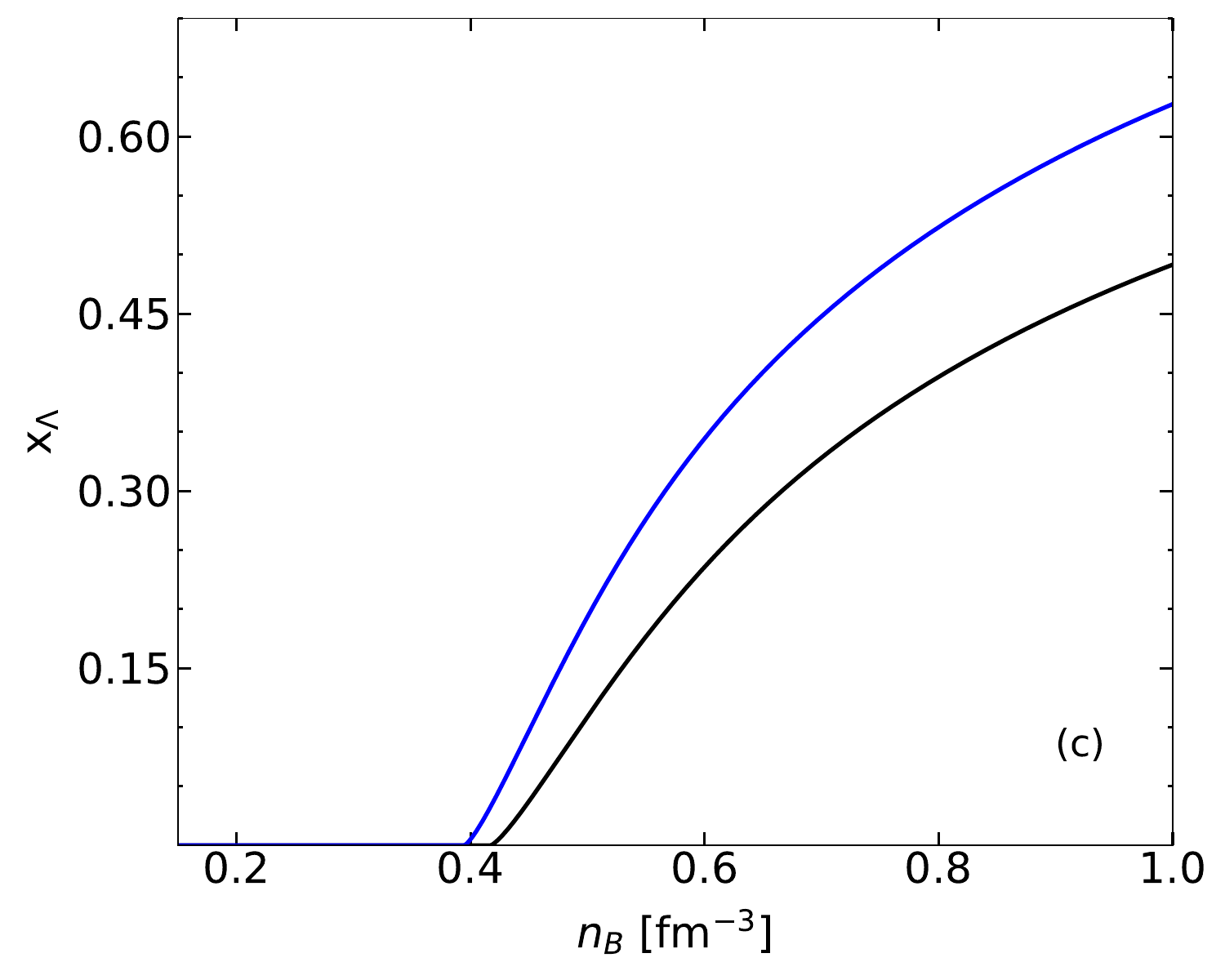} & \includegraphics[width=0.48\textwidth]{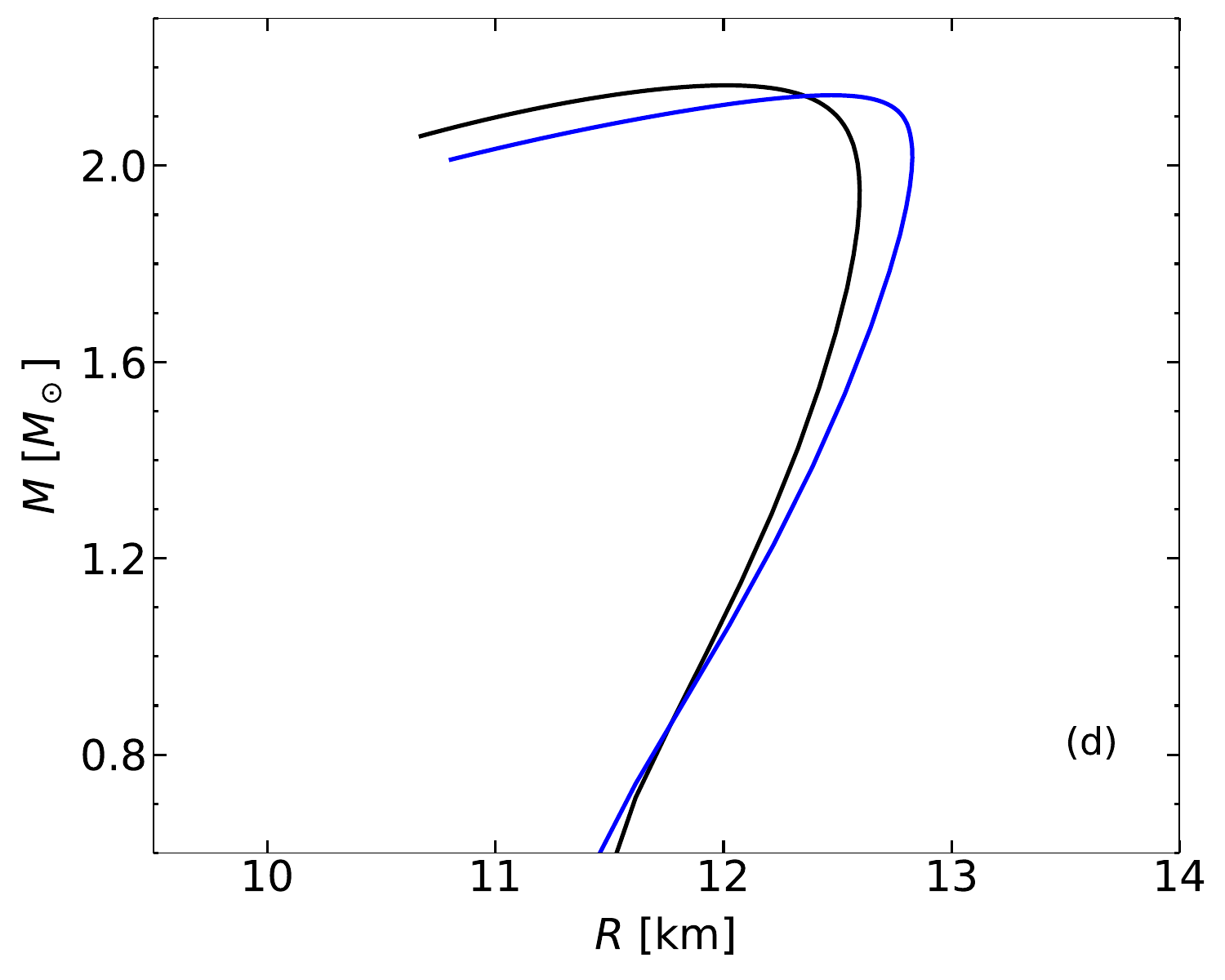} 
    \end{tabular}
   \caption{Pressure,  proton fraction, $\Lambda$ fraction as a function of baryon density $n_B$, and mass-radius for two most probable EOS models from the distributions for each cases, respectively, see text for details.}
    \label{fig:compare_eos}
\end{figure*}
\begin{table}
   \centering
    \begin{tabular}{c|c|c}
        \hline 
        \hline
        Parameters & Ratio &  SU(6)  \\
        \hline 
        \hline
        $a_\sigma$ & 8.1390878  &  8.455436  \\
        $b_\sigma$ & 2.8764833  &  2.5580113  \\
        $c_\sigma$ & 2.5879794  &  2.7895479  \\
        $d_\sigma$ & 4.7013457 & 4.8476099  \\
        $a_\omega$ & 10.260011 &  10.572641  \\
        $b_\omega$ & 2.1250266 &  2.101002  \\
        $c_\omega$ & 5.5274538 &  6.4384114  \\
        $d_\omega$ & 0.8638553 &  1.1297941  \\
        $a_\rho$ & 0.0467881  &  -0.52037038   \\
        $b_\rho$ & 6.3669419  &  6.6442192   \\
        $c_\rho$ &  0.4820617 &  0.4337751   \\
        $d_\rho$ & -0.5770346 &  -0.5955957  \\ 
        \hline
        $R_{\sigma\Lambda}$ & 0.6416294  & 0.618898 \\
        $R_{\omega \Lambda}$ & 0.6965185   &   \\
        $R_{\sigma* \Lambda}$ &  0.1824414  & 0.4657355  \\
        $R_{\phi \Lambda}$ & -0.5654369     &   \\    
        \hline
        $ n_{\mathrm{sat}}$ (fm$^{-3}$)  & 0.1695012   & 0.1674386  \\
        $m^*$ & 0.6712814   & 0.6599269  \\
         $ E_{\mathrm{sat}}$ (MeV) & -16.658733  & -16.80156  \\
         $ K_{\mathrm{sat}}$ (MeV) &  338.00469   & 347.68284 \\
         $ E_{\mathrm{sym}}$ (MeV) &  30.646875   & 29.21992  \\
         $ L_{\mathrm{sym}}$ (MeV) & 21.95607  & 22.21677  \\
        $K_{\mathrm{sym}}$ (MeV) &  -224.98675   & -214.53886  \\
        \hline
        $U_\Lambda^N(n_{\mathrm{sat}})$ (MeV) & -28.776765   & -28.546427  \\
        $n_\Lambda^{\mathrm{onset}}$ (fm$^{-3}$) & 0.4186679  & 0.3951551  \\
        \hline  
        \hline 
    \end{tabular}
    \caption{Parameters for the most probable models from the Ratio and SU(6) EOS ensembles, used as our example models together with the resulting values for the nuclear matter parameters.}
    \label{tab:compare_eos}
\end{table}

To perform a systematic study of $\Lambda$ hyperons in neutron star matter, we have generated
two different ensembles of EOSs, namely, Ratios, and $SU(6)$ (see Sect. \ref{sec:coupling_par}). 
The sample EOSs are not all equally probable, given that their maximum masses are affected by the hyperon content obtained in the individual realizations. Therefore, we performed a Bayesian analysis of all samples for the three cases by assigning to each of them a likelihood using different nuclear and astrophysical filters. The details of the procedure can be found in Refs. \cite{universe7100373,Mondal:2022cva,Char:2023fue,Scurto:2024ekq,Montefusco:2024xrx,Char:2025zdy}. In this section, we briefly summarize the steps of our analysis and discuss the differences with respect to the protocols employed in the papers cited above, and the extension to include the effects of the hyperons. Our analysis comprises multiple stages to use our computation resources efficiently. 
\begin{itemize}
    \item First, we generate a base set of samples for the nucleonic part of the EOS. We follow the framework developed in Ref. \cite{Char:2025zdy} to generate the parameters of the coupling functionals for the nucleonic EOS. We sample the parameters of the  nucleonic Lagrangian to calculate the corresponding NMPs. In this stage, we have used constraints from saturation properties of symmetric nuclear matter, and the $\chi$-EFT calculations of the PNM and SNM pressures as obtained by \textcite{Huth:2020ozf}, see their Figure 1. Following \cite{Scurto:2024ekq, Char:2025zdy}, we use a modified Gaussian distribution to implement the $\chi$-EFT constraints. Then, we used a Nested Sampling method to obtain equally weighted samples of model parameters using the {\tt PyMultiNest} software \cite{Buchner:2014nha}.
    \item Then, we impose additional constraints on the NMPs from AME2016 nuclear mass table \cite{Wang:2017fhd}. At this point, we denote this set of nucleonic model parameters as our prior set informed by theoretical and experimental nuclear physics. 
    \item Once we find an optimized set of nucleonic model parameters, we calculate 
the high density EOS in $\beta$-equilibrium incorporating hyperons. For each EOS within the nucleonic sample, we vary the hyperon coupling parameters for the two cases
Ratio and $SU(6)$ within the ranges given in Table \ref{tab:parameters}, and calculate
the hyperon potential ${\cal U}_\Lambda^N(n_{sat})$ for symmetric matter at saturation. We keep the hyperon parameters that generate ${\cal U}_\Lambda^N$ within $(-50,-10)$ MeV. We continue this process for  the entire nucleonic sample and augment one unique hyperon parameter set to each of the prior samples for  the two cases. Then, we continue to calculate the full high-density part of the EOS up to $\sim 7 n_{\mathrm{sat}}$.
    \item For the low-density part of the EOS, we have used the compressible liquid drop model developed by \textcite{Carreau:2019zdy}. This framework uses the NMPs corresponding to each of the parameter sets and uses a non-relativistic  \textbf{metamodel} (NRMM) to calculate the crust EOS. We joined the NRMM crust to the NRMM core at the crust-core transition. Then the low-density NRMM EOSs and the high-density relativistic EOSs containing hyperons are joined at the nuclear saturation density, exactly as in \cite{Char:2025zdy}. 
    \item At this stage of our analysis, we solve the TOV equations to calculate NS mass, radius, and tidal deformability sequences using the unified EOSs constructed in the previous step. Then, we impose the astrophysical constraints from pulsar mass observations and tidal deformability estimations from GW170817 \footnote{LVK collaboration,~\href{https://dcc.ligo.org/LIGO-P1800115/public}{https://dcc.ligo.org/LIGO-P1800115/public}}, following Refs. \cite{universe7100373,Mondal:2022cva,Char:2023fue,Scurto:2024ekq,Montefusco:2024xrx}. We have further applied a Gaussian constraint on  ${\cal U}_{\Lambda}^N(n_{\mathrm{sat}})$  with a mean of $-30$ MeV and standard deviation of $5$ MeV\footnote{A recent work on calculating the hyperon single-particle potential for infinite nuclear matter from chiral hyperon-nucleon interactions suggests a larger range for ${\cal U}_\Lambda^N(n_{\mathrm{sat}})$, with values approximately between $-25$ and $-60$ MeV \cite{Jinno:2025vgm}, thus in particular the possibility of a stronger attraction than assumed here. Studying the implications of these findings will be kept for future work.}. Even though for the prior we sampled ${\cal U}_\Lambda^N$ between -50 and -10 MeV, we applied this additional constraint to study the maximal effect of the hyperonic potential anticipating more precise experimental measurements in the future. Finally, we use Bayes theorem to compute the posterior distributions of the quantities of our interest. 
\end{itemize}

In summary, we have defined  nuclear physics informed priors that include 
constraints from $\chi$-EFT, AME2016, and the choice of nuclear matter parameter 
ranges used in our previous work \cite{Char:2023fue,Char:2025zdy}. 
 Our posteriors were determined after applying the astrophysical constraints and the requirement that the hyperon  potential in symmetric matter at saturation is compatible with the hypernuclear data, by using a Gaussian likelihood  ${\cal N}(-30,5)$ MeV. 
All in all, we have used $\sim 32000$ nucleonic parameter sets in our ``prior" 
and added $\Lambda$ hyperons following the two different ways as described in 
Sec \ref{sec:coupling_par}. For each of the nucleonic sets, we have added one unique 
instance of the two cases with the samples drawn uniformly from the parameter 
ranges in Tab. \ref{tab:parameters}. 

To compare the hyperonic and nucleonic hypotheses quantitatively, we estimate the Bayesian evidence of each ensemble as the astrophysical likelihood averaged over the shared nuclear physics-informed prior,
\begin{equation}
\begin{split}
\mathcal{Z} &\;=\; \frac{\sum_i w_i^{\mathrm{pr}}\,\mathcal{L}_i^{\mathrm{astro}}}
                        {\sum_i w_i^{\mathrm{pr}}}, \qquad
B = \mathcal{Z}_1/\mathcal{Z}_2 , \\[4pt]
\mathcal{L}_i^{\mathrm{astro}} &= P(\mathcal{C}_{M_{\mathrm{max}}}\,|\,X_i)\,
                                  P(\mathcal{C}_{\mathrm{LVK}}\,|\,X_i) ,
\end{split}
\end{equation}
where $w_i^{\mathrm{pr}}$ is the prior weight (AME2016) common to all ensembles and the sum runs over the parameter sets. Here, $P(\mathcal{C}_{M_{\mathrm{max}}}\,|\,X_i)$ is the likelihood of the maximum mass, taken as the Gaussian cumulative distribution evaluated at the EOS model's TOV maximum mass, and $P(\mathcal{C}_{\mathrm{LVK}}\,|\,X_i)$ is the GW170817 tidal deformability likelihood. Both are defined as in Refs. \cite{Char:2023fue,Char:2025zdy}. The Bayes factor between two hypotheses is the ratio of their evidences. Because the nucleonic prior and its weights are identical across ensembles, this ratio is insensitive to the overall normalization of the prior.

The astrophysical likelihood $\mathcal{L}^{\mathrm{astro}}$ deliberately excludes the hypernuclear constraint on $U_\Lambda$. In our treatment this constraint enters the hyperonic sector in two ways: a wide prior window, $-50 < U_\Lambda^N(n_{\mathrm{sat}}) < -10$~MeV, that bounds the hyperonic support, and a Gaussian likelihood $\mathcal{N}(-30, 5)$~MeV that shapes the posterior within it. The Gaussian factor is applied separately, so we omit it from the evidence ratio. The comparison then rests on the common astrophysical likelihood shared by all ensembles. The only residual hypernuclear conditioning is the wide window, which removes little coupling volume. The resulting Bayes factor therefore measures the discriminating power of the neutron-star observations alone. Contamination from the hypernuclear data, which constrain the hyperon couplings but have no nucleonic counterpart, is minimal. The extra hyperon coupling dimensions are nonetheless retained in the evidence and marginalized over their prior ranges. The comparison thus includes the corresponding Occam penalty for the added parametric freedom.

\section{results}\label{sec:results}
We discuss the results obtained in the present work with three distinctive foci
in mind. First, we concentrate only on the GDFM type functional form for the density-dependence Eq.(\ref{eq:GDFM})  and analyze  the impact of relaxing the $SU(6)$ hypothesis for
 incorporating hyperons in neutron star matter. Due to the restrictions posed in the hyperonic parameter space by the $SU(6)$ case, we anticipate higher dispersion in the predictions  in the Ratio case. 
We consider this latter prescription as the more realistic. Indeed, there is no reason why the predictions from the naive quark model should be respected by the baryonic couplings and the increased freedom in the parameter variation seems essential to simultaneously respect the attractive/repulsive character of the hyperonic interactions and verify the maximum mass constraint.

To evaluate the possible influence of hyperons in NSs, in the second subsection we compare the hyperonic Ratio case with the purely nucleonic case.  Since the nucleonic part of the functional remains exactly the same as in Ref. \cite{Char:2025zdy}, we do not discuss the isoscalar and isovector nuclear parameter distributions and their mutual correlations explicitly in the main text. They are provided in appendix \ref{appendix_nmps}.  Finally, the impact of hyperons obviously depends on the nucleonic EOS which is assumed, and misleading conclusions may be drawn if only a restricted set of nucleonic EOS is considered. It is therefore important to assess the model dependence of   different relativistic functional forms during the inclusion of hyperons in neutron star matter. To this aim, we compare the GDFM and TW models, albeit only with the $SU(6)$ case. Finally, we briefly discuss the thermodynamic stability of our models and the question of a phase transition at hyperon onset.

\subsection{Comparison among different ways to fix hyperon couplings}\label{sec:compare_hyp}
We start by showing in Fig. \ref{fig:compare_eos} the behavior of  example EOS models from each setting,  which are the most probable EOSs of their 
respective distributions. 
The parameters corresponding to these example EOS models 
are given in Tab. \ref{tab:compare_eos} and the data for both models will be made publicly available on the \textsc{CompOSE} data base~\cite{Typel:2013rza,CompOSECoreTeam:2022ddl}. 
In panel (a), (b) and (c) of Fig. \ref{fig:compare_eos}, the behavior of pressure, proton fraction 
and $\Lambda$-hyperon fraction are plotted as function of baryon number density, respectively. 
In panel (d) the  mass-radius relations are displayed. What one can clearly appreciate  from the figure is that even though they represent the most probable of their respective distributions, subject to the same constraints, their overall behaviors are different. In particular, their nucleonic parts are not the same: to meet the NS maximum mass constraint, the $SU(6)$ one has a stiffer nucleonic part and higher proton fraction at low densities, which decreases at higher densities after the onset of hyperons. Correspondingly, the predicted radii for intermediate mass NS are relatively high. This result is in qualitative agreement with previous studies \cite{Malik:2022jqc}.
However, it is interesting to observe that this strong correlation between the presence of hyperons and the global characteristics of the star fades away when the simplified $SU(6)$ restriction is relaxed. Indeed, no apparent softening is observed with the representative  Ratio EOS, in spite of the fact that  all EOS produce  similar hyperon fractions and TOV maximum mass. In the rest of the subsection, this statement will be further confirmed and quantified by looking at the different EOS models and stellar observables  with the complete statistical samples.

\begin{figure}
    \centering
    \includegraphics[width=0.5\textwidth]{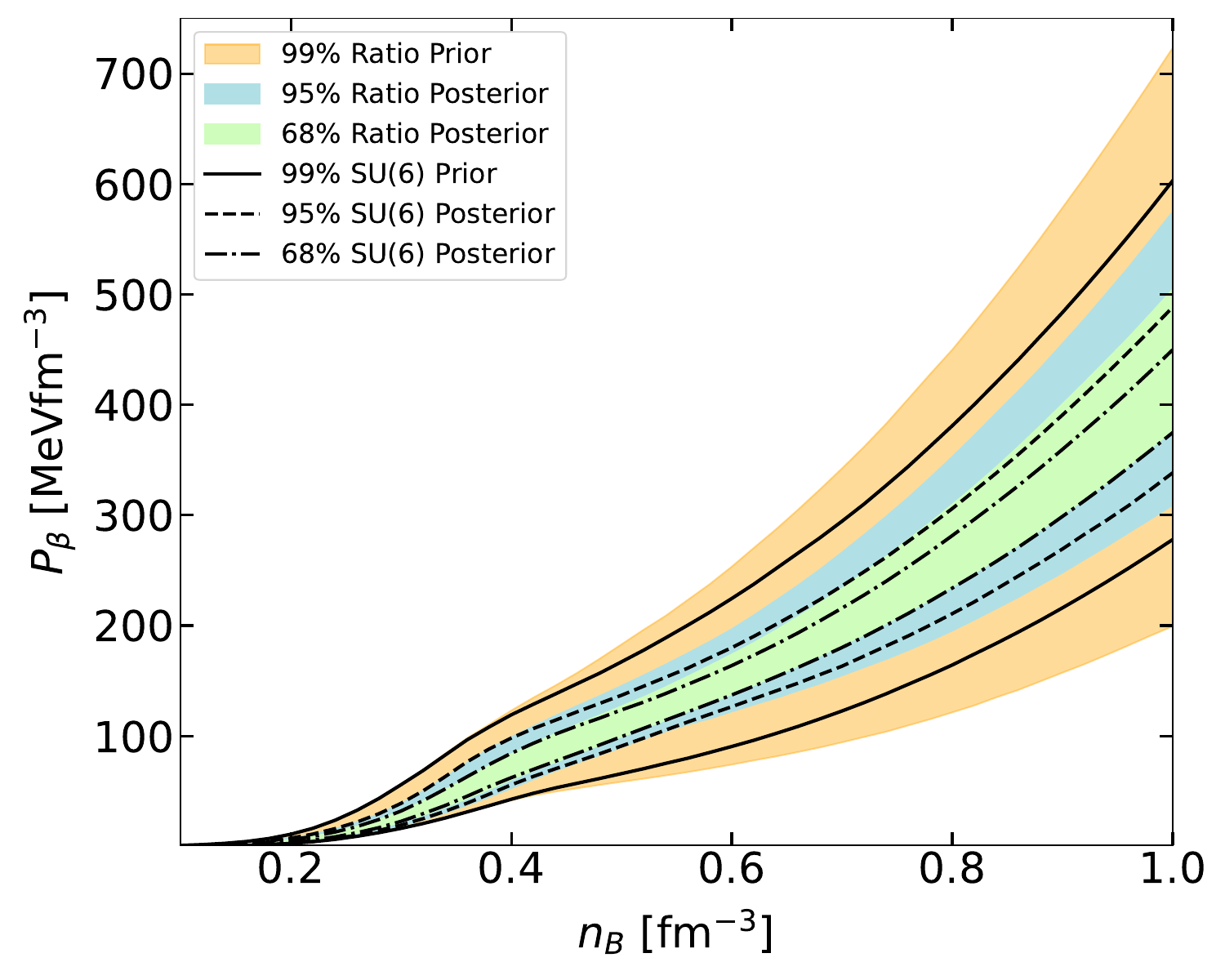}
\caption{Contours of pressure at $\beta$-equilibrium as function of baryon number density at different CI for the  Ratio and $SU(6)$ cases. }
    \label{fig:p-rho}
\end{figure}

In Fig. \ref{fig:p-rho}, we have shown the 68\% and 95\% credible intervals (CI) of the EOS posteriors along with their 99\% prior contours for the  Ratio and SU(6) cases.  We have used color-filled regions for the Ratio setting using  light blue for the 95\% posterior, light green for the 68\% posterior and light orange for the 99\% prior, respectively. For the  SU(6) setting, we have used black lines. We have used different linestyles, solid, dashed, and dashdot for 99\% prior, 95\% and 68\% posteriors, respectively. We have followed this convention in all other contour plots in this subsection. Though nucleons are always dominant in the star with respect to $\Lambda$-hyperons (see Figure \ref{fig:sf} below), the hypotheses made on the hyperon couplings have an important effect not only on the average behaviors, as already seen in Figure \ref{fig:compare_eos}, but also on the dispersion of the predictions. In particular,
 we can see that the Ratio prior provides  a larger variation in the EOS space due to its larger modeling freedom. The SU(6) prior contours are smaller  due to its more restrictive parameter space. The trend continues to the posterior contours as well. The Ratio produces larger posterior than SU(6). In any case, both extremely soft and extremely stiff EOSs are ruled out by the combined effect of the NS maximum mass and tidal deformability constraints. 

\begin{figure}
    \centering
    \includegraphics[width=0.5\textwidth]{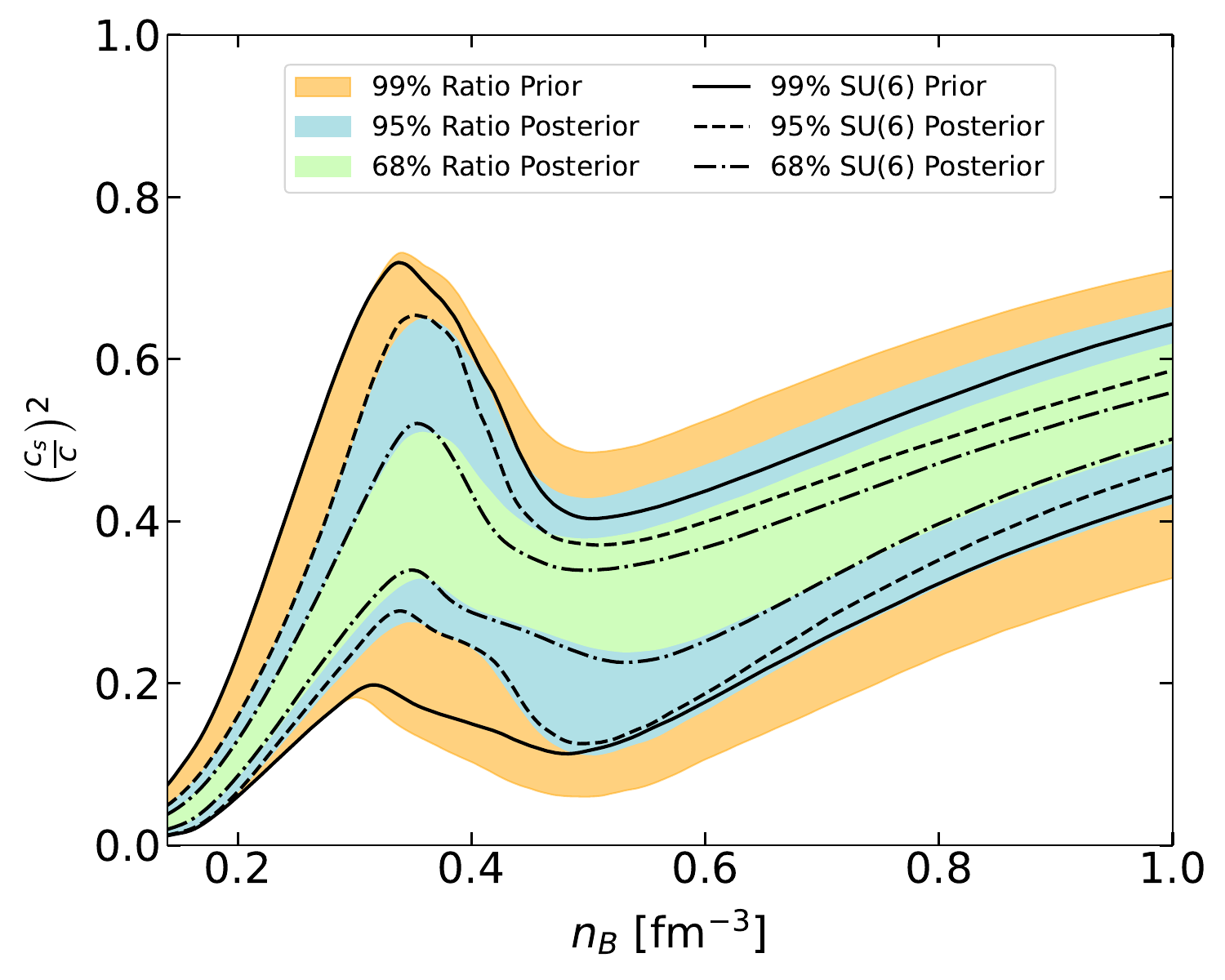}
\caption{Same as Fig. \ref{fig:p-rho}, but for speed of sound.}
    \label{fig:cs}
\end{figure}
The larger dispersion in the distribution for the Ratio setting can be seen in the speed-of-sound posteriors, too, see Fig.~\ref{fig:cs}. The prominent decrease in speed of sound around $n_B \sim 0.3$ fm$^{-3}$ is thereby the imprint of the onset of hyperons and the corresponding softening of the EOS.
\begin{figure}
    \centering
    \includegraphics[width=0.5\textwidth]{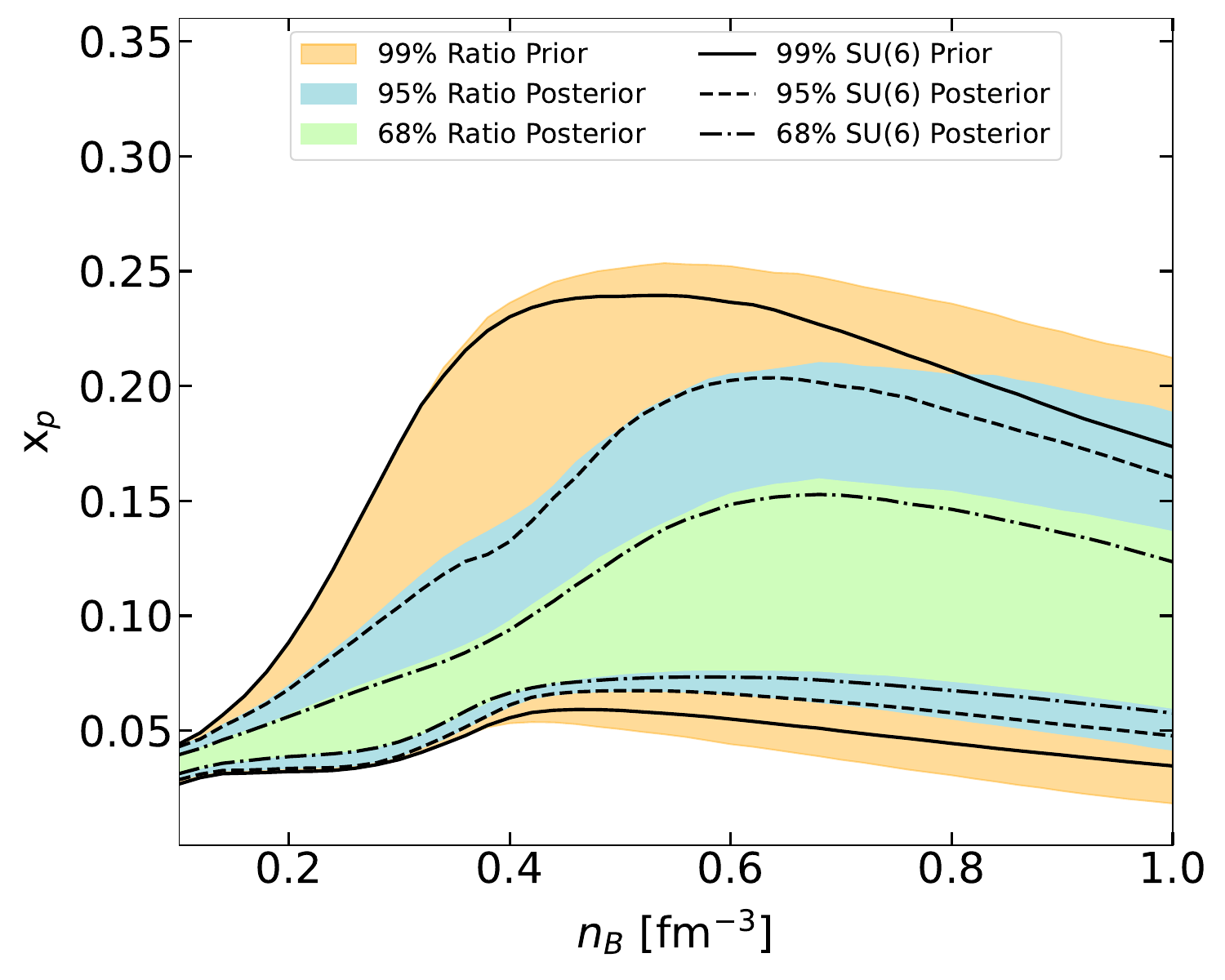}
\caption{Same as Fig. \ref{fig:p-rho}, but for proton fraction.}
    \label{fig:xp}
\end{figure}

In Fig. \ref{fig:xp}, we show the proton fraction contours as a function of baryon number density corresponding to the EOS contours in Fig. \ref{fig:p-rho}. Please keep in mind that although $\Lambda$ hyperons can well mimic the overall effect of hyperons on the EoS and the total hyperon fraction inside a NS, the proton fraction is very sensitive to the potential presence of charged hyperons and the values in the above figure should thus be regarded with some care. We can, however, still note some trends for the comparison of the Ratio and $SU(6)$ settings. In both cases, the proton fraction is reduced due to hyperon onset with a larger dispersion for Ratio as expected. In particular at high densities larger proton fractions can be reached for Ratio. As a general statement, we can see that the flexibility of the GDFM functional form allows exploring a large set of proton fraction with respect to more restrictive choices for the nucleonic couplings, in agreement with previous studies \cite{Char:2023fue,Char:2025zdy,Scurto:2024ekq,2025arXiv250318889S}, see also Section~\ref{sec:compare_su6}.

\begin{figure}
    \centering
    \includegraphics[width=0.5\textwidth]{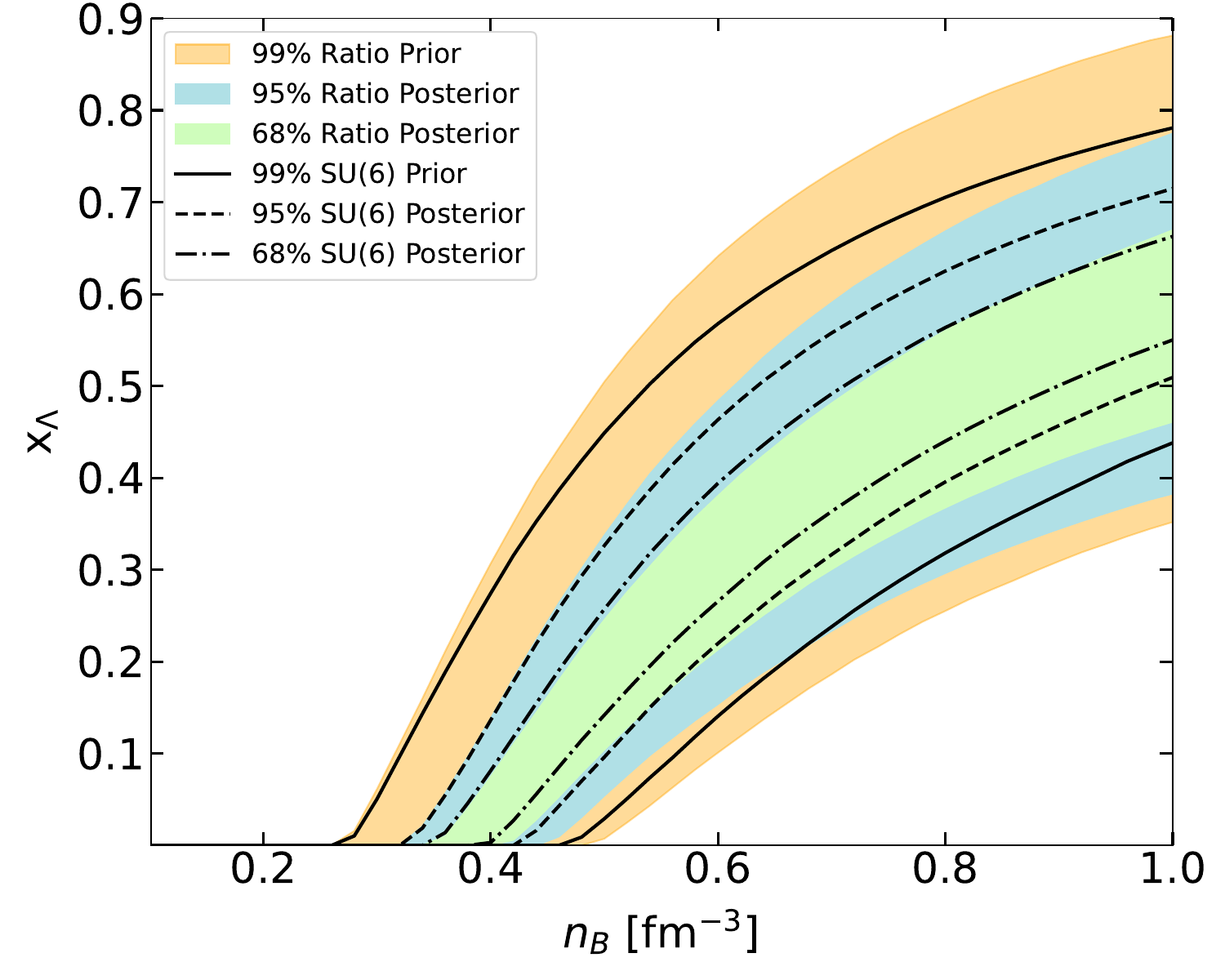}
\caption{Same as Fig. \ref{fig:p-rho}, but for $\Lambda$ fraction.}
    \label{fig:xla}
\end{figure}

The hyperon fractions for  both settings are shown in Fig.~\ref{fig:xla}. The higher proton fractions found for Ratio at high densities are consistent with the lower hyperon fractions explored by Ratio at the same densities , as well as the  behavior of the EOS model distributions in Figs. \ref{fig:p-rho} and \ref{fig:cs}.

\begin{figure}
    \centering
    \includegraphics[width=0.5\textwidth]{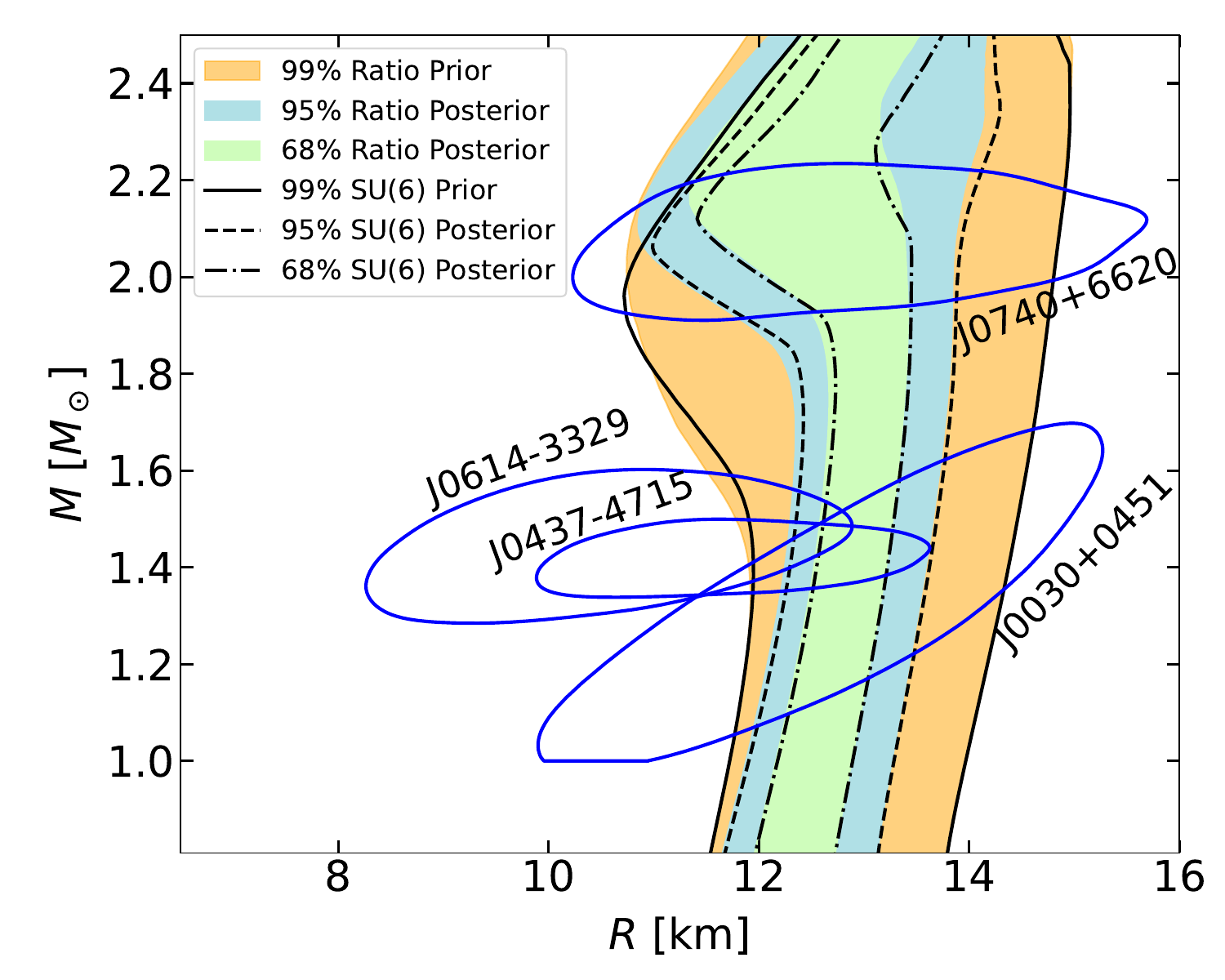}
\caption{Contours of mass-radius relations at different CIs corresponding to the EOS models shown in Fig.~\ref{fig:p-rho}. The solid blue contours represents the 95\% CI of the different NICER sources \cite{Riley:2019yda,Riley:2021pdl,Choudhury:2024xbk} (see text for details).}
    \label{fig:mr}
\end{figure}

Now, let us in turn examine the global stellar properties arising from the different settings. 
We have reported the M-R contours for the two cases in Fig.~\ref{fig:mr}. We take a fixed grid of mass points and find the radius distribution at those mass points, corresponding to the EOS contours in Fig. \ref{fig:p-rho}. We see that  for lower mass stars ($M \lesssim 1.4 M_{\odot}$) the contours are almost overlapping.  The central densities of these low mass stars are either below hyperon threshold or hyperon concentrations are very low, thus this region is  dominated by the underlying  nucleonic models. We thus do not observe any considerable impact of the different treatment of hyperonic couplings and the way the NS maximum mass constraint is fulfilled on the posterior distribution of nucleonic models. This feature starts to change around $\sim 1.6 M_\odot$ and above. We find the prior boundaries of Ratio extended to slightly smaller radii.  Upon applying the astrophysical constraints, the softest EOS models  are excluded and the allowed region starts with larger radii also for low-mass stars. The stiffest EOS models are, as expected, equally excluded, and very large radii become disfavored. Let us stress that altogether, the EOS models with hyperons can still produce high maximum masses and reasonably small radii. The posterior distributions of both settings become very similar and only at the highest masses with $M \gtrsim 2.1 M_{\odot}$ the larger parametric freedom of the Ratio setting leads to a distribution reaching slightly smaller radii. This is perfectly consistent with all our findings on the EOS models and reflects mainly the possibility of Ratio to have additional repulsion in the hyperonic couplings at high densities leading to smaller hyperon fractions, larger proton fractions and stiffer EOS. We have also plotted the $95\%$ contours of the simultaneous mass-radius observations from NICER and found that our M-R contours are consistent with those observations.

\begin{figure}
    \centering
    \includegraphics[width=0.5\textwidth]{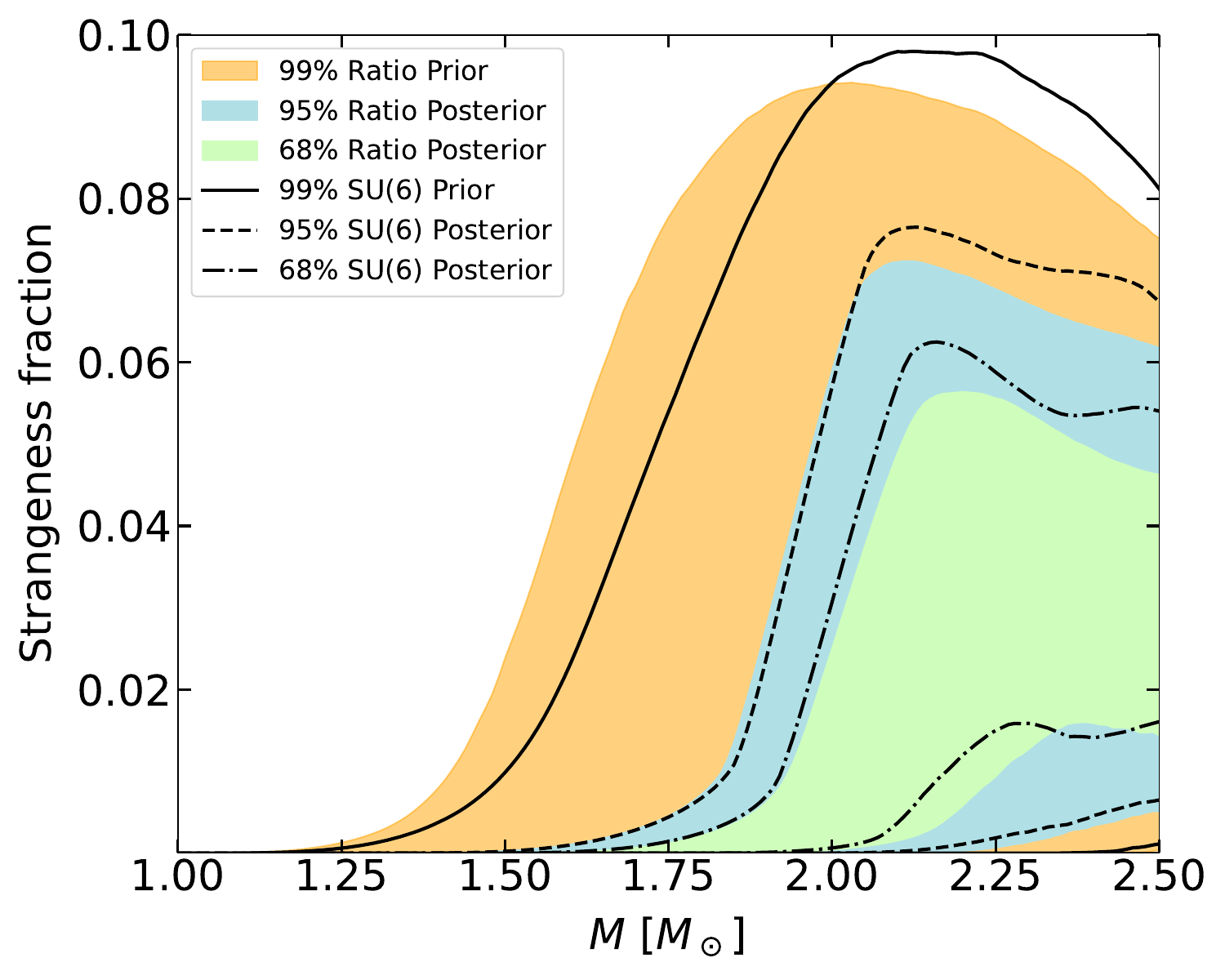}
\caption{Total strangeness fractions as a function of NS mass corresponding to the EOS model ranges shown in Fig.~\ref{fig:p-rho}}
    \label{fig:sf}
\end{figure}

Average values of different observables together with their 90\% confidence interval are reported in Table \ref{tab:ns_params}. Though some systematic trends are observed in the averages, the  distributions are largely compatible within the uncertainties.
All in all, we can say that our very limited theoretical knowledge of the values and density behaviors of the hyperon couplings does not induce important ambiguities or biases in the predictions of the NS mass-radius relation. The same is not fully true concerning the stellar composition and the maximum mass, as we now show.

In Fig. \ref{fig:sf}, we have displayed the total strangeness fraction  as a function of the NS mass. 
To find the strangeness fraction, we first calculate the total number of baryons ($A$) and hyperons ($A_{\Lambda}$) in a star as
\begin{eqnarray}
    A &=& 4 \pi \int_0^R \left(1 - \frac{2M(r)}{r} \right)^{-1/2} r^2 n(r) dr, \nonumber\\
    A_{\Lambda} &=& 4 \pi \int_0^R \left(1 - \frac{2M(r)}{r} \right)^{-1/2} r^2 n_{\Lambda}(r) dr.
    \label{eq:sf}
    \end{eqnarray}
The total strangeness fraction in a star is then defined by $A_{\Lambda}/3A$~\cite{Weissenborn:2011ut,Oertel:2014qza}. 

Not surprisingly, we find that the Ratio distribution is larger and in particular covers lower strangeness fractions for high-mass stars. This is consistent with the behavior of $x_\Lambda$ from Fig. \ref{fig:xla} where we found the possibility of lower  concentration of $\Lambda$'s at high densities for Ratio.

\begin{figure}
    \centering
    \includegraphics[width=0.5\textwidth]{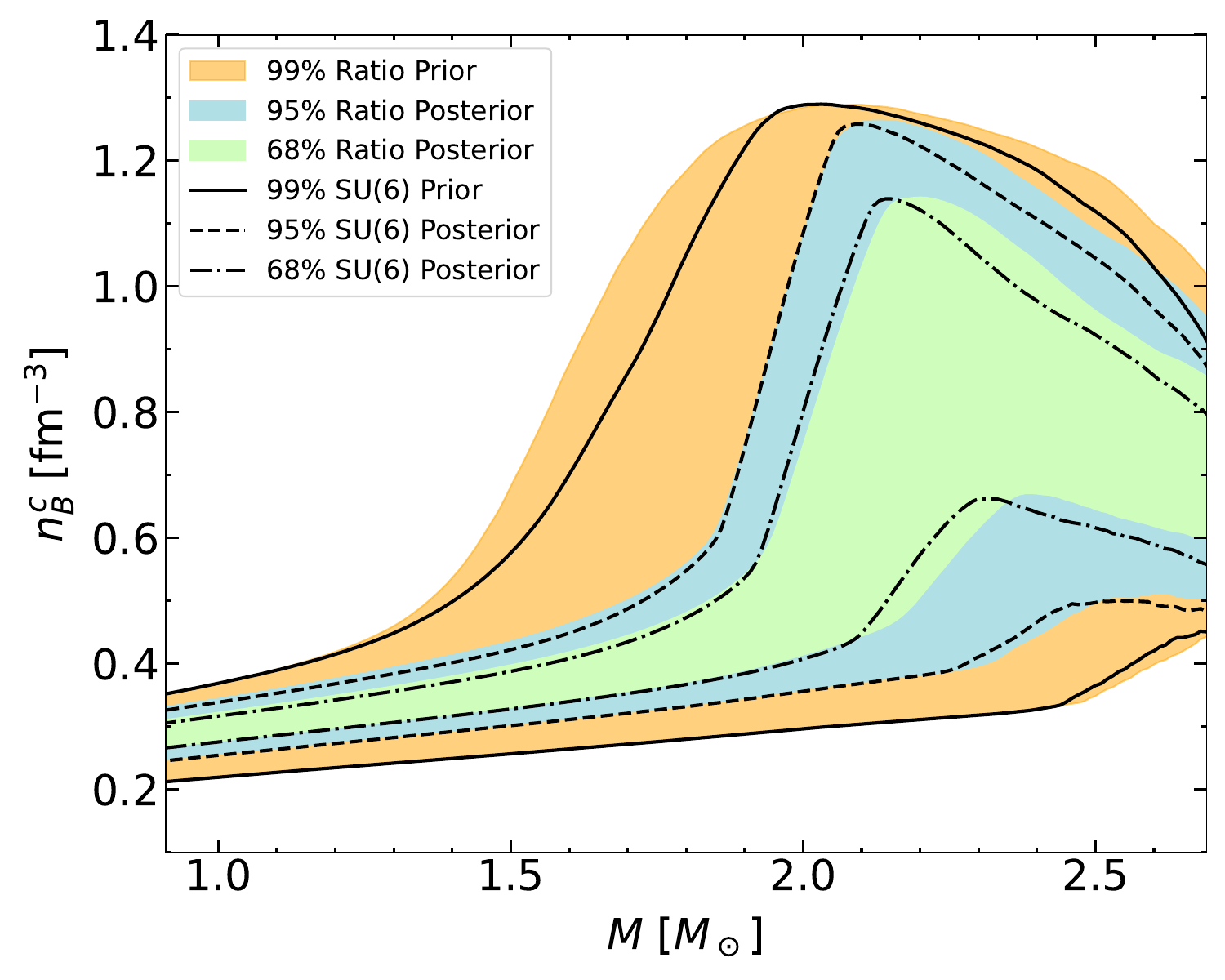}
\caption{Contours of central baryon number densities as function of NS mass corresponding to the EOS model ranges shown in Fig.~\ref{fig:p-rho}.}
    \label{fig:nbm_hyp}
\end{figure}

\begin{table*}[]
    \centering
    \begin{tabular}{c|c|c|c|c|c|c}
    \hline
    \hline 
           & $R_{1.4}$ (km) & $n_{B,c}^{1.4}$ (fm$^{-3}$) & $R_{2.0}$ (km) & $n_{B,c}^{2.0}$ (fm$^{-3}$) & $M_{max} (M_\odot)$ & $n_{B,c}^{M_{max}}$ (fm$^{-3}$) \\
    \hline 
    \hline 
     Ratio   & $12.89^{+0.55}_{-0.55}$ & $0.344^{+0.052}_{-0.040}$ & $12.95^{+0.72}_{-1.66}$ & $0.509^{+0.636}_{-0.135}$& $2.171^{+0.220}_{-0.149}$ & $0.827^{+0.162}_{-0.166}$ \\
     SU(6)   & $12.92^{+0.54}_{-0.55}$ & $0.339^{+0.049}_{-0.037}$ & $13.03^{+0.67}_{-2.01}$ & $0.499^{+0.696}_{-0.129}$ & $2.148^{+0.208}_{-0.135}$ & $0.809^{+0.155}_{-0.149}$  \\
    \hline 
    \hline
    \end{tabular}
    \caption{Median values of various NS quantities along with their 90\% CI values obtained from the Ratio and SU(6) posteriors, respectively. }
    \label{tab:ns_params}
\end{table*}

The maximum mass peaks well above the $2M_\odot$ limit. As seen from Fig. \ref{fig:mr}, we find that -although the differences are small, the Ratio distribution peaks at a slightly higher maximum mass. This again can be understood from the possibility of slightly lower hyperon fraction at high densities associated to this setting. Correspondingly,  the central density Ratio  also peaks at a slightly higher central density than SU(6). To assess the evolution of central densities with different masses, in Fig.~\ref{fig:nbm_hyp} we plot the 68\% and the 95\% posterior and 99\% prior distributions of the central densities as a function of NS mass. Overall, the ranges of central densities in the posteriors are very similar, except again at the highest masses, where the Ratio distribution is slightly larger allowing for slightly higher central densities in agreement with the results discussed before.

\subsection{Comparison between nucleonic and hyperonic matter within the GDFM model}\label{sec:compare_nuchyp}

\begin{figure*}
    \centering
    \begin{tabular}{cc}
        \includegraphics[width=0.48\textwidth]{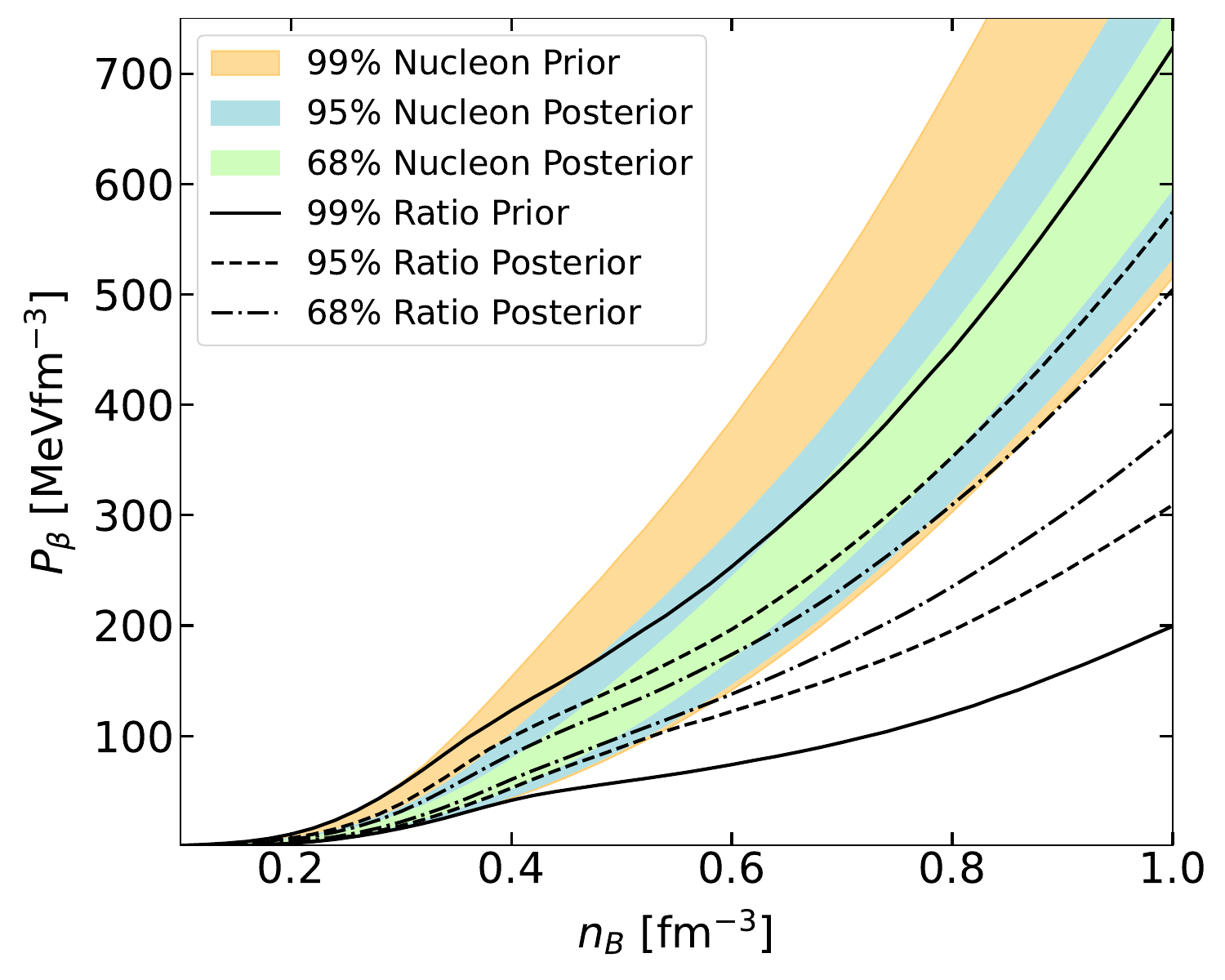} &  \includegraphics[width=0.48\textwidth]{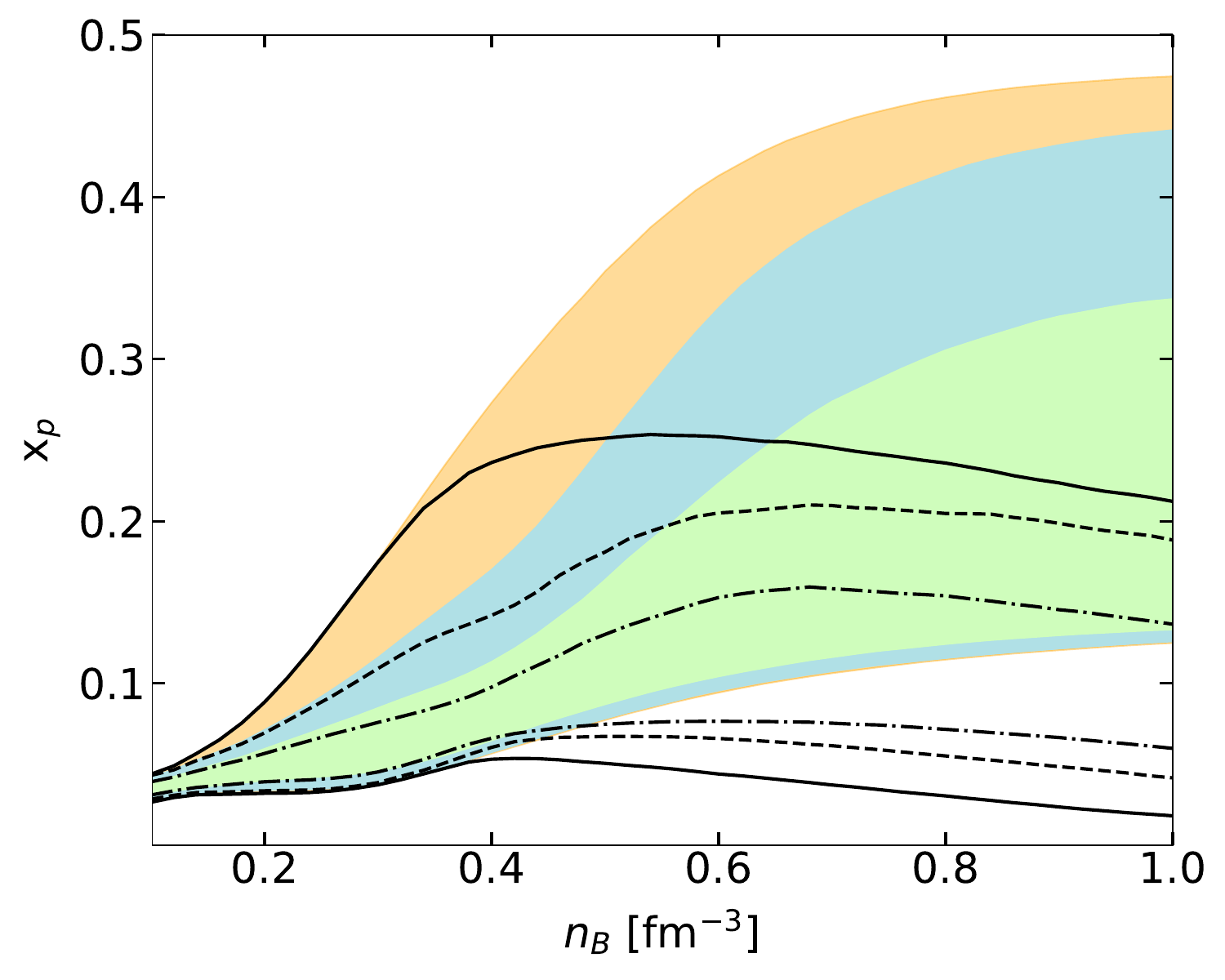} \\
	    \includegraphics[width=0.48\textwidth]{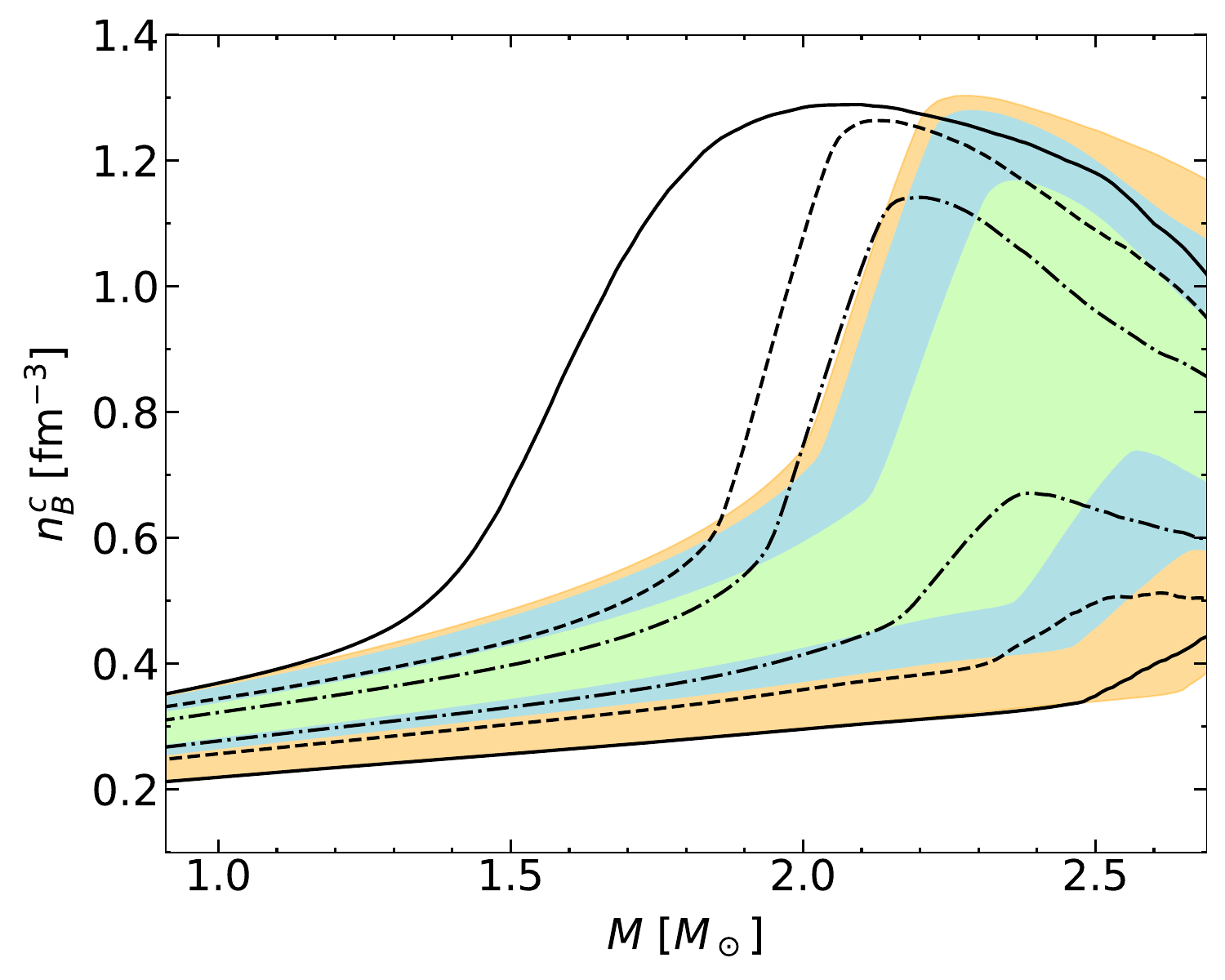} & \includegraphics[width=0.48\textwidth]{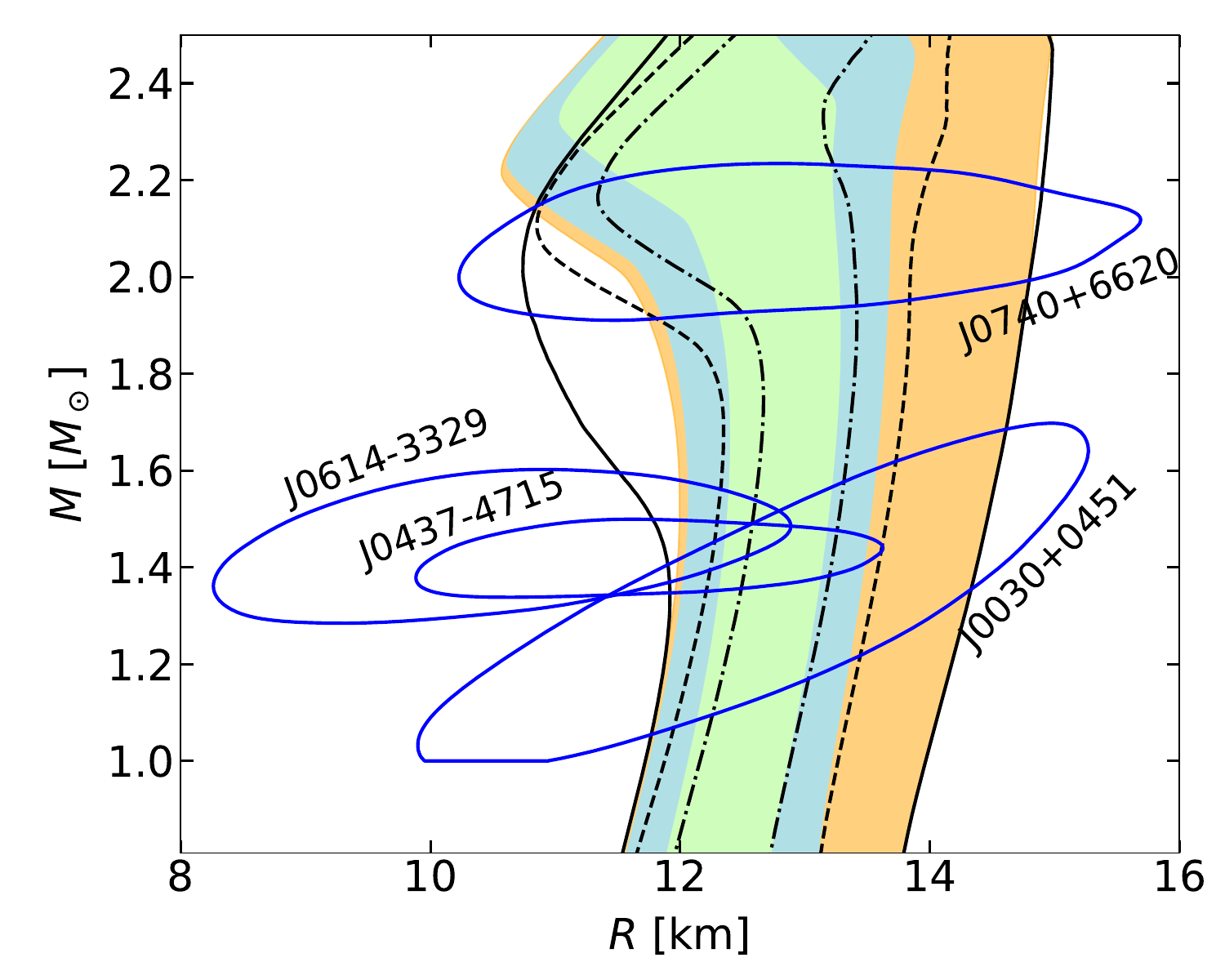} 
    \end{tabular}
   \caption{Pressure,  proton fraction as a function of baryon density ($n_B$) at $\beta$-equilibrium, central densities as a function of stellar mass, and mass-radius at different CI for the GDFM nucleonic and  Ratio hyperonic models. The solid blue contours represents the 95\% CI of the different NICER sources \cite{Riley:2019yda,Riley:2021pdl,Choudhury:2024xbk}.}
    \label{fig:compare_nuchyp}    
\end{figure*}

In Figure \ref{fig:compare_nuchyp}, we compare distributions of nucleonic and hyperonic EOS models. We use the Ratio setting for this purpose as it has the highest freedom in the hyperon sector, meaning that the probability of a bias in our results due to unjustified hypotheses on the hyperonic interactions will be minimized. The underlying GDFM nucleonic model has been already explored in detail in Refs. \cite{Char:2023fue, Char:2025zdy,Scurto:2024ekq}. We see in the upper left panel of Figure \ref{fig:compare_nuchyp} a clear softening due to hyperons after the onset. This leads to a substantially lower proton fraction as shown in the upper right panel of the same figure, although the range of the proton fractions remains quite wide similar to the nucleonic case. 
In particular, the threshold value for a possible stellar fast cooling via the direct nucleonic Urca process $x_p\approx 1/9$ \cite{2006PhRvC..74c5802K} is still met for the heaviest stars, in addition to possible hyperonic Urca processes. A sizeable, even if smaller effect is also seen in the central density of the heaviest stars (lower left panel of Figure \ref{fig:compare_nuchyp}), the central densities for hyperonic stars being slightly higher than their purely nucleonic counterparts. As can be seen, this is mainly an effect of the reduced maximum mass due to the presence of hyperons.

These results are in qualitative agreement with previous studies \cite{Sun:2022yor,Malik:2022jqc,Providencia:2023rxc, Huang:2024rvj}. However, these studies reported a sizeable effect of the presence of hyperons in the 
prediction of the NS radius, though with somewhat contradictory results (the effect of allowing for hyperons in the EOS leads to a radius increase in the results of Ref.\cite{Malik:2022jqc,Providencia:2023rxc} even before hyperon onset due to the interplay between hyperonic softening of the EOS and the NS maximum mass constraint, while a decrease is observed in the work of \cite{Sun:2022yor}).
Conversely, the effect of hyperons is  almost negligible in our study as shown in the 
  mass-radius relation displayed in the lower right panel of Figure \ref{fig:compare_nuchyp}.
Specifically, the average $R_{1.4}$ for  for Ratio is 12.89 km and for SU(6) is 12.92 km, with an increase  only on the percent level  with respect to its nucleonic value of $\sim 12.72$ km reported in \cite{Char:2023fue,Char:2025zdy}. From the statistical distribution, it seems thus extremely tricky to distinguish between hyperonic and nucleonic EOS  just by observing mass and radius. Let us, however, point out that the precise observation of NS masses and radii can still reveal the presence of hyperons via the specific behavior of the M-R relation at hyperon onset~\cite{Bauswein:2025dfg}.

We can make this statement quantitative by computing the Bayesian evidence for each ensemble under the astrophysical likelihood (maximum mass and tidal deformability), estimated as the likelihood averaged over the shared nuclear physics-informed prior [see Sec. \ref{sec:bayes}]. The resulting Bayes factors between the hyperonic and nucleonic models are $\ln B = -0.89 \pm 0.02$ for the Ratio case and $\ln B = -0.96 \pm 0.01$ for SU(6). On the Jeffreys scale as revised by \textcite{Kass:1995loi}, a magnitude $|\ln B| < 1$ is ``not worth more than a bare mention.'' The astrophysical data therefore express no significant preference between hyperonic and nucleonic descriptions, quantitatively confirming that the two hypotheses cannot be discriminated on the basis of current bulk observations. We stress that this reflects the constraining power of present data rather than an intrinsic equivalence of the models. A genuine discrimination would require either substantially more precise mass-radius measurements or composition-sensitive observables. Among the latter, the curvature of the M-R relation at hyperon onset~\cite{Bauswein:2025dfg}, the direct Urca threshold and the  associated cooling behavior \cite{2006PhRvC..74c5802K} can  signal the presence of hyperons. The underlying composition change driving these signatures is the total strangeness fraction, which becomes non-zero above $\sim 2\, M_\odot$, displayed in Fig. \ref{fig:sf}.

Still, some differences appear for higher mass stars above $2M_\odot$, which reflect the non-zero strangeness fraction of the star, when hyperons set in. As expected from the increased EOS softness, the median of the distribution of maximum masses for the ensemble of sequences remains around $\sim 2.17 M_\odot$ (see, table \ref{tab:ns_params})  with the inclusion of hyperons, where for nucleons only it was found around  $\sim 2.37 M_\odot$ \cite{Char:2023fue,Char:2025zdy}.

%%%%%%%%%%%%%%%%%%%%%%%%%%%%%%%%%%%%%%%%%%%%%%%%%%%%
\subsection{Comparison of implementation of hyperons with different underlying nucleonic models}
\label{sec:compare_su6}

\begin{figure*}
    \centering
    \begin{tabular}{cc}
        \includegraphics[width=0.48\textwidth]{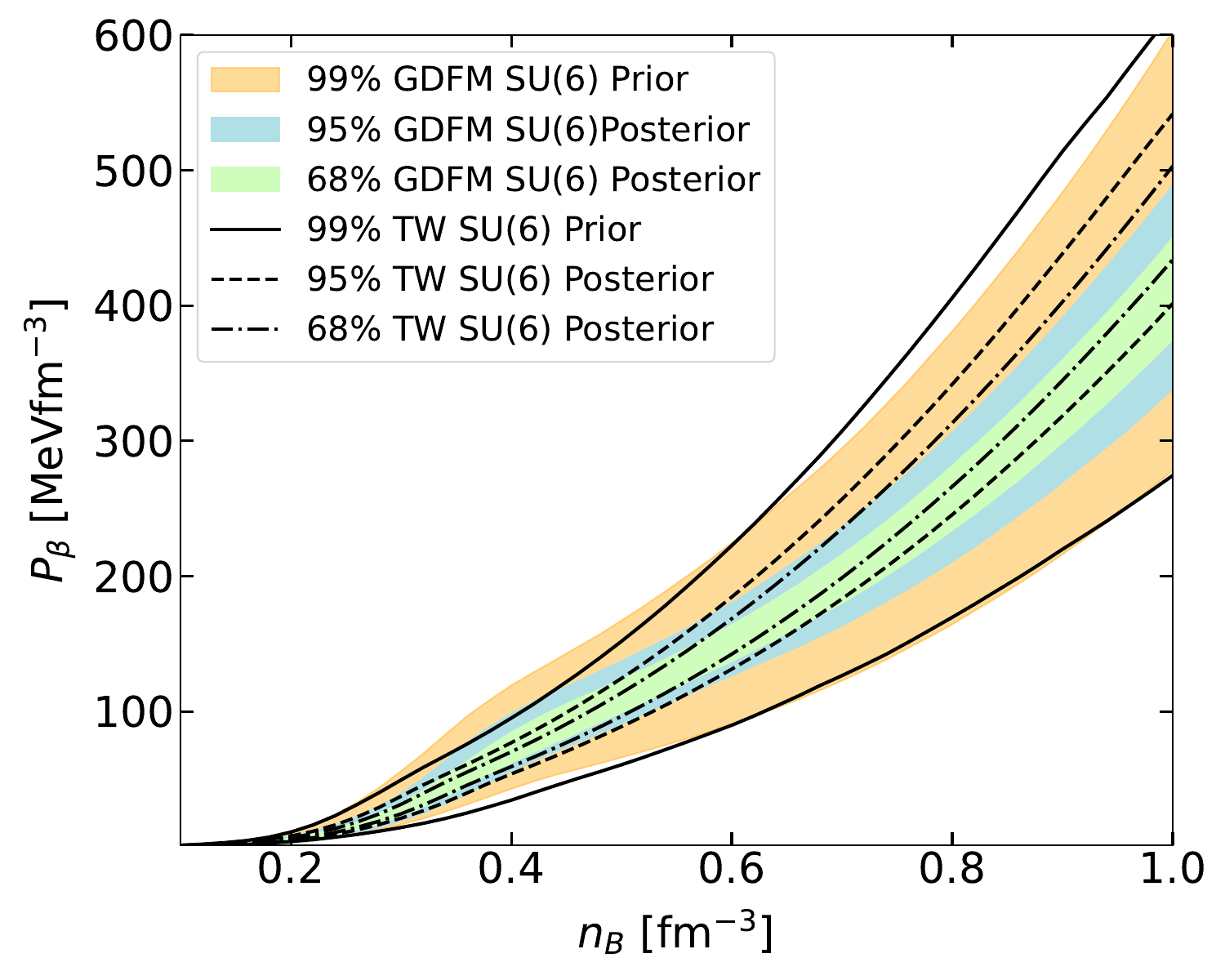} &  \includegraphics[width=0.48\textwidth]{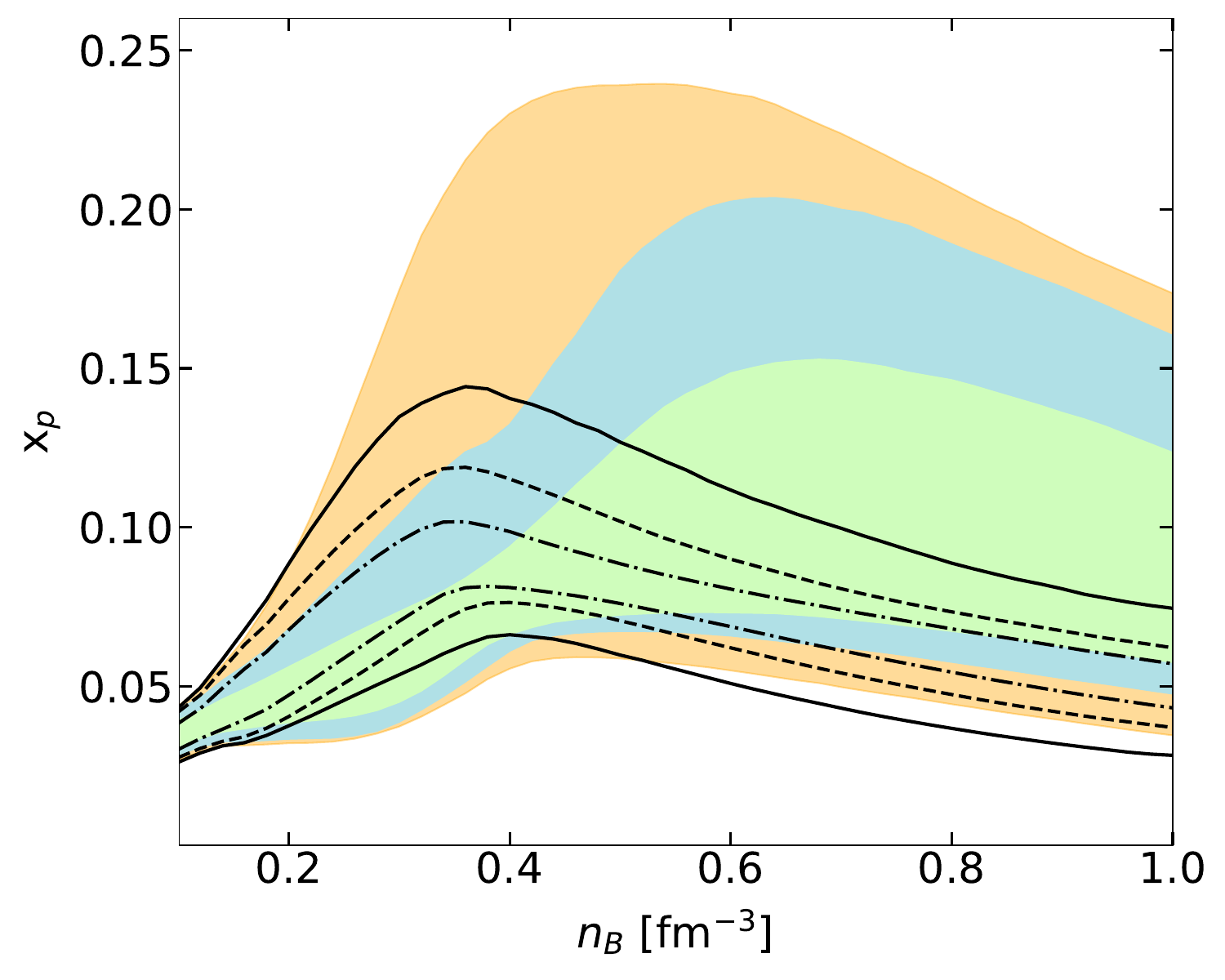} \\
	    \includegraphics[width=0.48\textwidth]{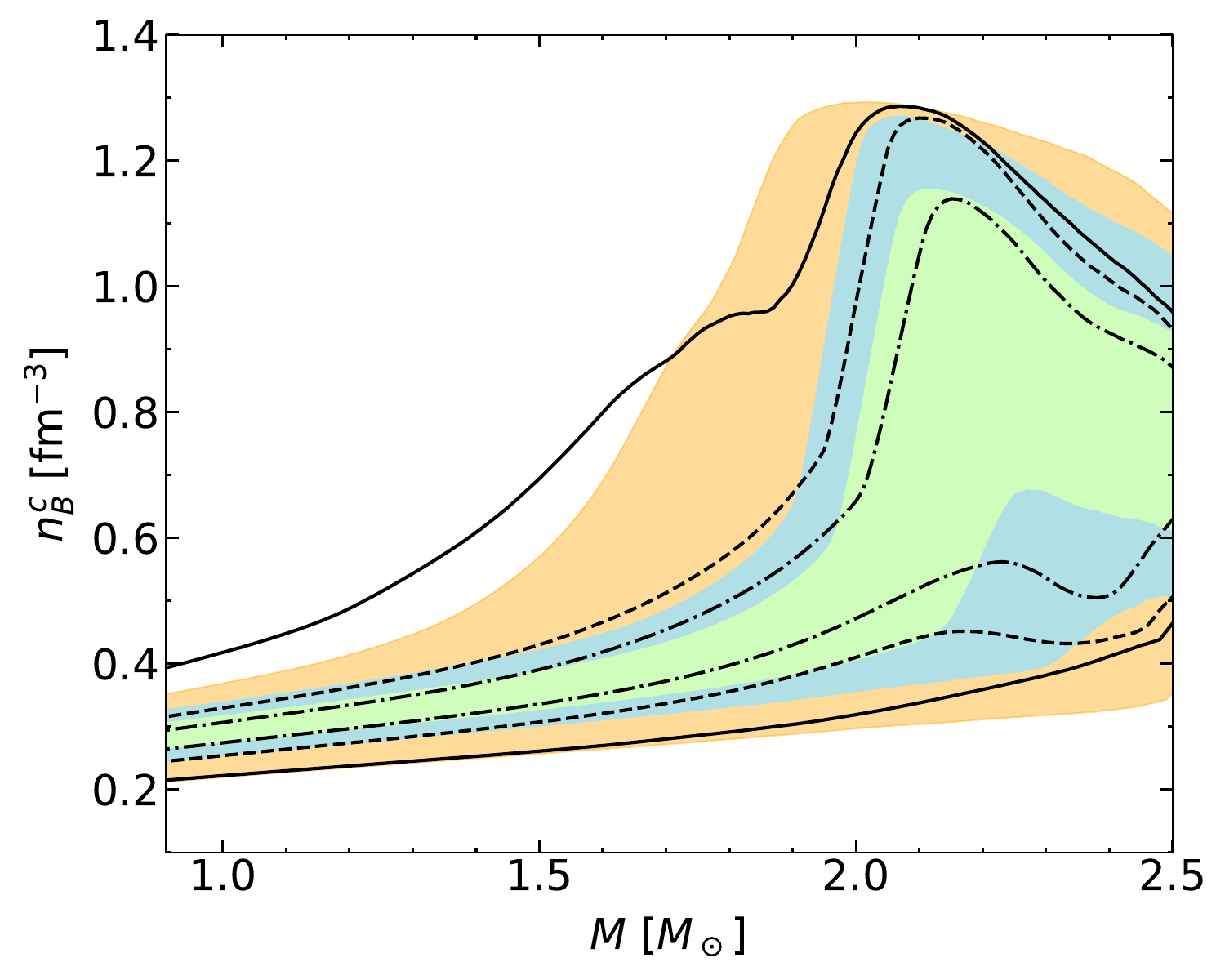} & \includegraphics[width=0.48\textwidth]{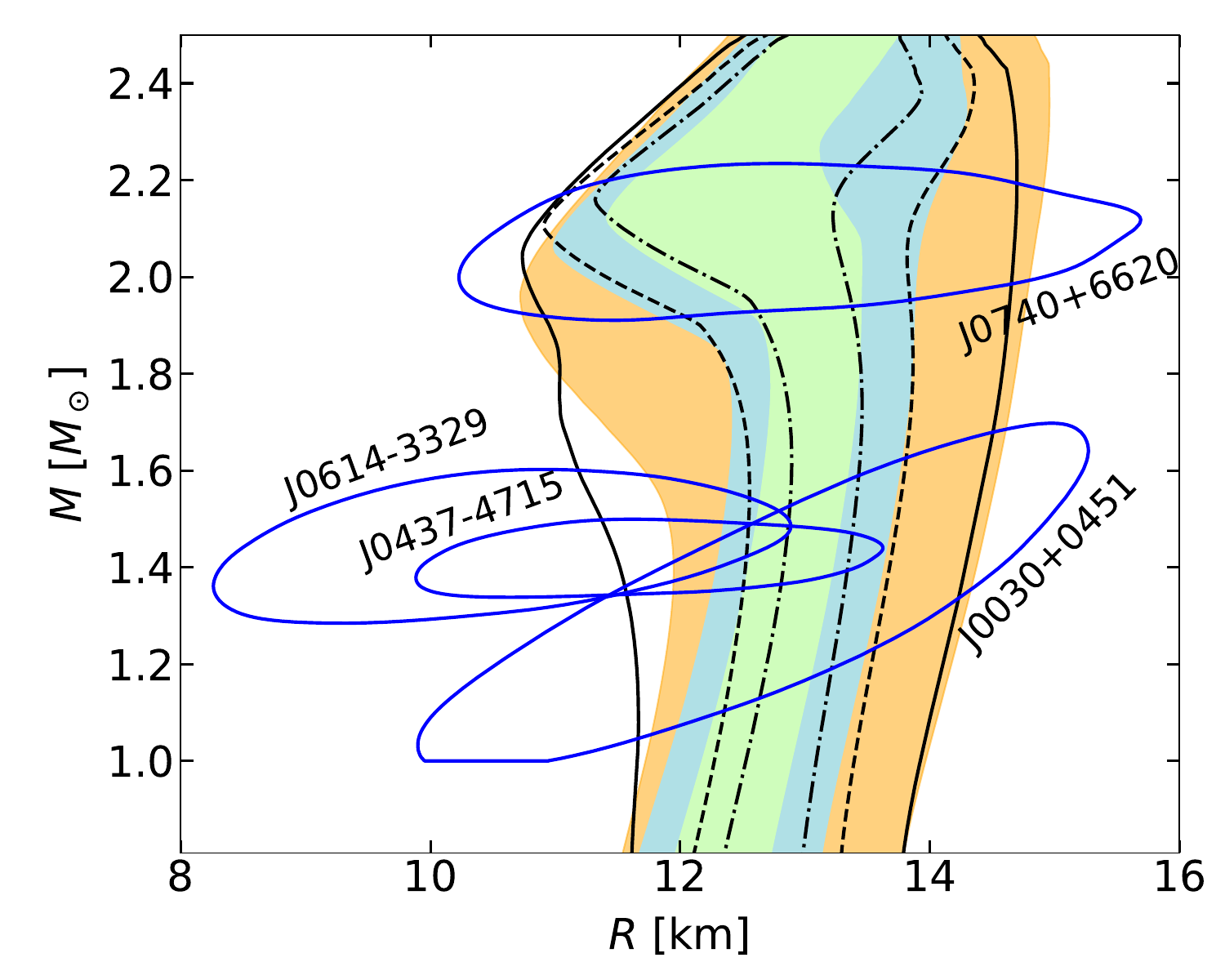} 
    \end{tabular}
   \caption{Pressure,  proton fraction as a function of baryon density ($n_B$) at $\beta$-equilibrium, central densities as a function of stellar mass, and mass-radius at different CI for the SU(6) cases with GDFM and TW nucleonic models. The solid blue contours represents the 95\% CI of the different NICER sources \cite{Riley:2019yda,Riley:2021pdl,Choudhury:2024xbk}}
    \label{fig:compare_su6}    
\end{figure*}

Though it is well recognized that realistic relativistic density functional models, in the absence of non-linear terms, must consider  density-dependent coupling constants, the functional form of this density-dependence is not well settled and different choices might induce  biases in the model predictions.  In order to check the possible model dependence due to the functional form of the density-dependence of the couplings, and additionally compare our results to the findings of Refs.\cite{Malik:2022jqc,Providencia:2023rxc}, we have decided to implement the hyperons using one of their models, namely the TW \cite{Typel:1999yq} density-dependence, given by Eq.(\ref{eq:gamadefault}). We have constructed full distribution of unified EOS models with TW nucleonic models with hyperons consistent with the  different settings for the treatment of the hyperon couplings described previously, and performed a Bayesian analysis. For details of the implementation of the nucleonic models, see Ref.~\cite{Char:2025zdy}. Concerning the hyperonic couplings,  it is particularly important to note that, in order to allow a meaningful comparison between the TW and GDFM functionals, we have used the ranges of $R_{\sigma\Lambda}$ and $R_{\sigma *\Lambda}$  given in Table \ref{tab:parameters}.  This is different from the setup of Ref.~\cite{Malik:2022jqc}, where the authors did not include the $\sigma*$ meson, and used a much narrower  $R_{\sigma\Lambda}$ range of $(0.609,0.622)$ leading to a very narrow distribution in the ${\cal U}_{\Lambda}^N(n_{\mathrm{sat}})$ values. In Figure \ref{fig:compare_su6}, we show the comparison between the results within the two functionals for the SU(6) coupling scheme, which is the choice in Refs.
\cite{Malik:2022jqc,Providencia:2023rxc}. Though we have varied the TW parameters over a range sufficiently large to insure comparable domains for the EOS priors, we can see that the the TW functional form leads to slightly softer EOS at lower densities, with an increased stiffening at higher density to support the maximum mass constraint.  
Deviations are also seen in the central densities (lower left panel) and, most interestingly, in the mass-radius relation displayed in the lower right panel of the figure. Specifically, the 68\%  and 95\% level for the mass-radius of TW is smaller than GDFM up to $2M_\odot$ and the stars with masses lower than $2M_\odot$ consistently produce larger radii for TW. For the TW SU(6), we find the median of $R_{1.4}$ to be $\sim 13.1$ km, while it stays at $\sim 12.92$ km for GDFM SU(6). For reference, the nucleonic TW has $R_{1.4} \sim 12.89$km and for GDFM, it is $\sim 12.72$ km. This shows that the larger radii for intermediate mass stars within TW are already present at the nucleonic level,  and the increase after including hyperons is is on the percent level for both TW and for GDFM. As discussed above, the main difference between our study and previous literature~\cite{Malik:2022jqc,Providencia:2023rxc} is the increased freedom in the scalar sector for the hyperonic couplings which as a consequence do not require stiffer EOS before hyperon onset to comply with the NS maximum mass constraint and thus do not lead to the prominent shift to larger radii for intermediate mass stars observed in Refs.~\cite{Malik:2022jqc,Providencia:2023rxc}.  The median of the distribution of $M_{max}$, in case of TW, reduces from $\sim 2.42 M_\odot$ for the nucleonic case to $\sim 2.16 M_\odot$ after incorporating the hyperons within the SU(6). For the case of GDFM SU(6), it decreases from $\sim 2.37 M_\odot$ from nucleonic to $\sim 2.15 M_\odot$ for the hyperons, very similar to the TW case.  
The most important difference is seen in the proton fraction (upper right panel of Figure \ref{fig:compare_su6}), which is much lower for the TW case, independently of  the presence of hyperons at a given density. This is the typical behavior of TW compared to GDFM, already discussed in Ref. \cite{Char:2025zdy}.

 As already discussed in \cite{Char:2025zdy}, the main reason for the differences observed between the functional dependence of TW and GDFM is due to the restricted parameter space in the isospin sector associated with the TW parametrization, as one can see comparing Eq.(\ref{eq:GDFM}) with Eq.(\ref{eq:TWrho}). The extra deviations between the results of this subsection and those reported in Refs.\cite{Malik:2022jqc,Providencia:2023rxc} can be ascribed to the absence of $\sigma^*$ couplings and the narrower ranges of $R_{\sigma\Lambda}$  explored in these previous works. Taken together, these comparisons reveal a clear hierarchy in the model dependence induced by the functional form. The EOS distribution and the global stellar properties, in particular the M-R relation and $R_{1.4}$, differ only modestly between GDFM and TW. The differences that do appear are already present at the nucleonic level rather than being introduced by the hyperons. The composition, by contrast, and specifically the proton fraction, is markedly more sensitive to the choice of functional form. This is the same pattern established for the purely nucleonic case in Refs. \cite{Scurto:2024ekq,Char:2025zdy}, where EOSs and M-R sequences from different functionals were found to be similar while their underlying proton fractions differed substantially. The functional form dependence in our hyperonic results thus mainly affects the weakly constrained composition sector, leaving the EoS and global NS observables comparatively robust.

%%%%%%%%%%%%%%%%%%%%%%%%%
\subsection{Stability} 
\begin{figure}
    \centering
    \includegraphics[width=0.5\textwidth]{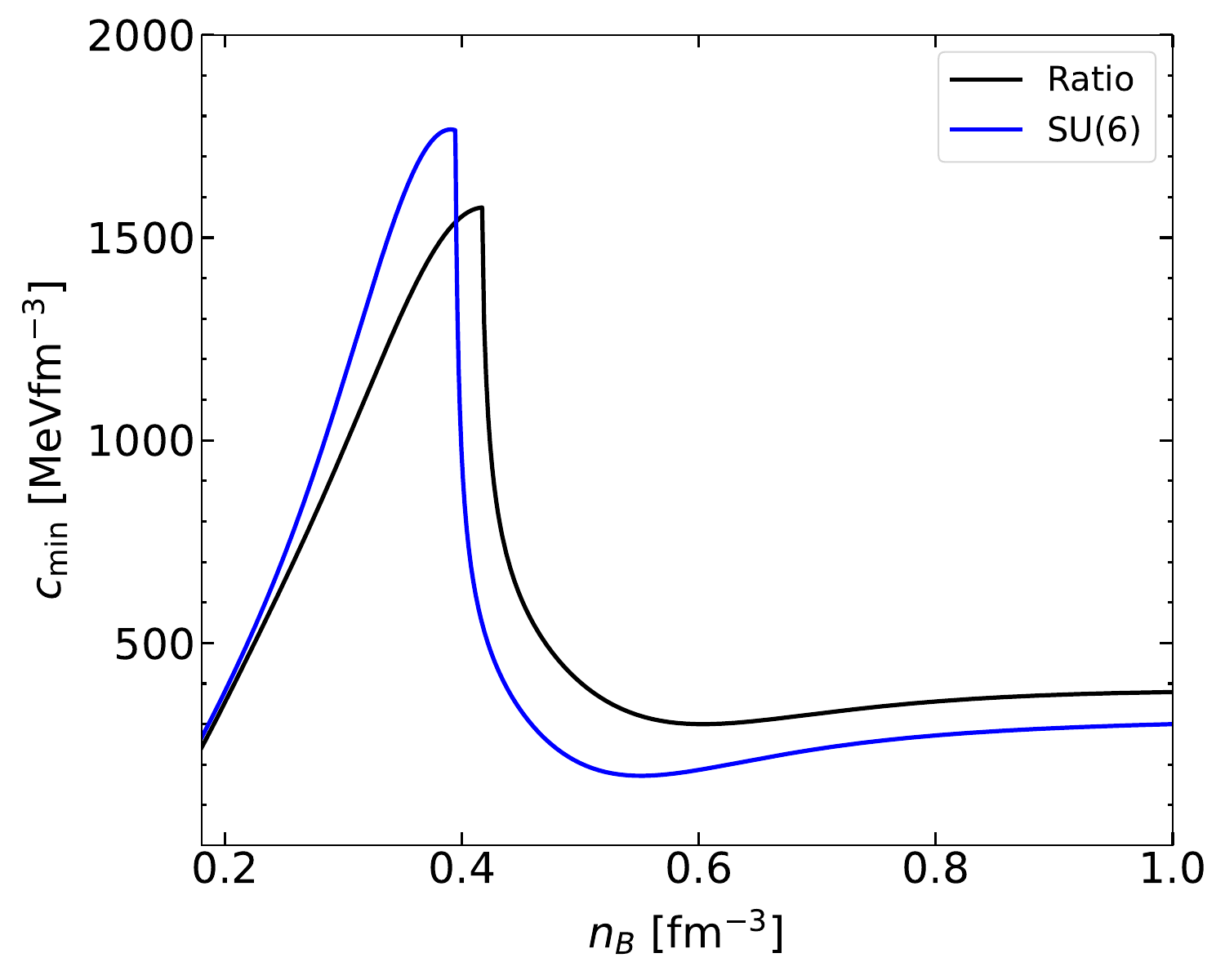}
\caption{Smallest eigenvalue of the curvature matrix of the energy density as a function of baryon density for the  example EOS models of Fig. \ref{fig:compare_eos}.}
    \label{fig:cmin}
\end{figure}

\begin{figure}
    \centering
    \includegraphics[width=0.55\textwidth]{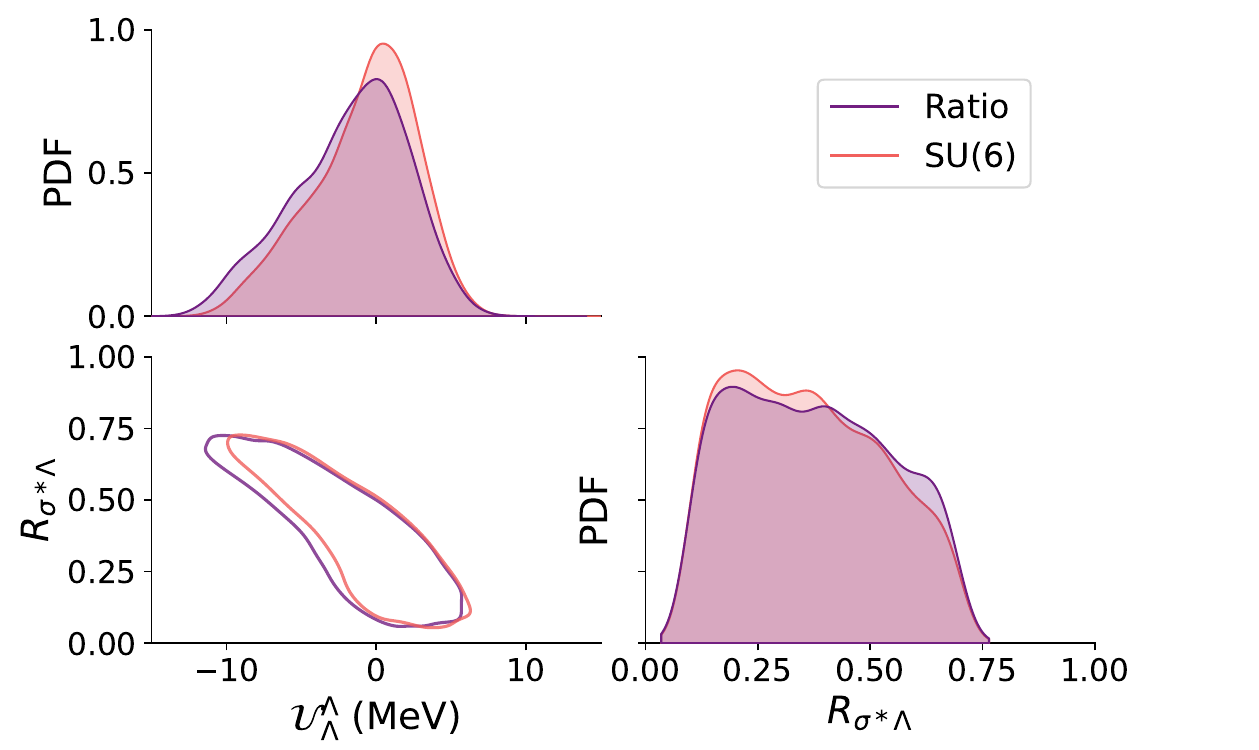}
\caption{Posterior distribution of  ${\cal U}_{\Lambda}^{(\Lambda)}  (n_{sat}/5)$ and $R_{\sigma* \Lambda}$, predicted from our calculations.}
    \label{fig:ull_post}
\end{figure}

Next, we discuss the stability of our hyperon EOSs. In figure \ref{fig:cmin}, the minimum eigenvalues, $c_{min}$, of the curvature matrices corresponding to the most favored example cases of  Ratio and SU(6) from table \ref{tab:compare_eos} are shown as functions of density. The kinks denote the onset of hyperons. None of our example models shows any negative eigenvalues, confirming the stability of these models. Then, we vary the $R_{\sigma* \Lambda}$  to determine whether we can find a region of the parameter space where such instabilities may arise. We can see that  for high enough values of $R_{\sigma* \Lambda}$, the EOS models indeed show instabilities. However, for all example models a value of $R_{\sigma^*\Lambda} \gtrsim 1$ is required for the instability to set in leading to ${\cal U}_{\Lambda}^{\Lambda}(n_{\mathrm{sat}}/5) \lesssim -20 $MeV.  Thus, the instability only appears if the attractive $\Lambda\Lambda$ coupling  overcomes the corresponding nucleonic coupling, which does not seem realistic \cite{Glendenning:1991es}. Also, experimentally, ${\cal U}_{\Lambda}^{(\Lambda)} (n_{\mathrm{sat}}/5)$ is expected to be around $\sim -5$ MeV \cite{Oertel:2014qza}. However, this phenomenological value should be regarded with some care due to the complexities in interpreting the experimental data. Instead of imposing this as a strict constraint, we have evaluated it a posteriori following the prescription from \textcite{Oertel:2014qza}. We have shown the posterior prediction between  ${\cal U}_{\Lambda}^{(\Lambda)}$ and $R_{\sigma^*\Lambda}$ in figure \ref{fig:ull_post}. Although the value of ${\cal U}_{\Lambda}^{(\Lambda)}$ is very uncertain since we extract a mean field potential in homogeneous matter from a measurement of the structure of a He hypernucleus, we still find it reasonably well reproduced and in a stable region. Therefore, the instabilities are far away from our region of interest. In conclusion, our chosen coupling ranges in table \ref{tab:parameters} always produce stable EOS models.

%%%%%%%%%%%%%%%%%%%%%%%%%%%%%%%%%%%%%%%%%%%%%%%%%%%%%
%%%%%%%%%%%%%%%%%%%%%%%%%%%%%%%%%%%%%%%%%%%%%%%%%%%%
\section{Conclusion}\label{sec:conclusion}
In this work, we have presented a generalized framework for incorporating hyperons within a relativistic approach. We have explored  two different approaches for incorporating the hyperonic couplings, the SU(6) approach, commonly used in the literature, and the ``Ratio'' one. We have performed Bayesian analyses following these approaches to understand how their parametric freedom influences the  distributions of EOS models and NS global quantities. For our posteriors, we have imposed the different constraints: $\chi$-EFT computations of pure neutron matter, AME2016 nuclear mass table, and a hyperon optical potential of ${\cal N}(-30,5)$ MeV , the maximum TOV mass  from pulsar mass measurements, and tidal deformability information from GW170817. We have found important overlaps for the two settings in the mass-radius diagram, although their underlying particle fractions can be quite different. We have found that the Ratio setting provides  slightly larger ranges for  various quantities of interest  than the SU(6) one. This follows from the larger inherent freedom relaxing the SU(6) hypothesis.  We have also compared the  Ratio hyperonic EOS models with their nucleonic GDFM counterparts. The  EOS clearly softens after the emergence of hyperons. But the onset density can vary over a large range of densities, thereby creating several EOS instances with varying hyperon fractions at neutron star densities. As a result, we see mass radius sequences with hyperons yet producing high maximum mass thereby providing the possibilities of smaller radii stars. This is one of the most important results that has not been explored previously. Finally, we have compared the effect of hyperon implementation with different underlying nucleonic density functionals to understand our results in light of previous works in the literature. We have seen that both --the choice of hyperon parameter ranges and the freedom of the underlying nucleonic density functionals-- govern the ranges of radii for intermediate stars, represented quantitatively here by $R_{1.4}$, found in the literature. With our choices, we have managed to generate smaller $R_{1.4}$ and consequently models with better consistency with the GW170817 data.

In the present work, we have restricted ourselves to the $\Lambda$ hyperon, for which the hyperon-nucleon interaction is the most tightly constrained by hypernuclear data. It is nonetheless worth commenting on what the inclusion of the heavier $\Sigma$ and $\Xi$ hyperons would imply. A repulsive $\Sigma$-nucleon potential, as favored by most current data, would suppress or delay the appearance of $\Sigma^-$, leaving the $\Lambda$ as the dominant strange species. A moderately attractive $\Xi^-$ potential would allow $\Xi^-$ to appear at high density and contribute additional softening of the EOS, with a corresponding further reduction of the maximum mass. These effects would sharpen the maximum mass tension and, within our framework, favor more repulsive high-density couplings, the same mechanism that we already find for the $\Lambda$. The central qualitative conclusions of this work, however, are not tied to a specific hyperon species. They can be summarized as (i) the radius shift observed when SU(6) couplings are imposed  disappears when this restrictive hypotheses is relaxed; (ii)  present astrophysical observations are virtually insensitive to the composition of matter. These behavior would be reinforced by introducing additional, even more weakly constrained, hyperonic couplings. We note that the existing Bayesian RMF studies including the heavier hyperons treat their couplings under more restrictive assumptions than we apply to the $\Lambda$.  \textcite{Sun:2022yor} infers only the $\Lambda$ couplings while holding the $\Sigma$ and $\Xi$ couplings fixed at empirical values, whereas \textcite{Huang:2024rvj} includes the $\Sigma$ and $\Xi$ scalar couplings but under tight, empirically centered Gaussian priors and within a single nucleonic functional family. A full multi-strange exploration under the relaxed coupling, functional agnostic philosophy adopted here is therefore a natural and
still open extension, which we leave to future work.

\section*{Acknowledgments}
This project has received funding from the European Union’s Horizon 2020 research and innovation programme under the Marie Skłodowska-Curie grant agreement No. 101034371. PC acknowledges the support from the European Union's HORIZON MSCA-2022-PF-01-01 Programme under Grant Agreement No. 101109652, project ProMatEx-NS. CM acknowledges partial support from the Fonds de la Recherche Scientifique (FNRS, Belgium) and the Research Foundation Flanders (FWO, Belgium) under the EOS Project nr O022818F and O000422. FG and MO acknowledge financial support from the Agence Nationale de la Recherche (ANR) under the contract ANR-22-CE31-0001-01. FG acknowledges partial support from the In2p3 Master project MAC.

\appendix \section{Nuclear matter parameters}\label{appendix_nmps}

\begin{figure*}
    \centering
\includegraphics[width=0.9\textwidth]{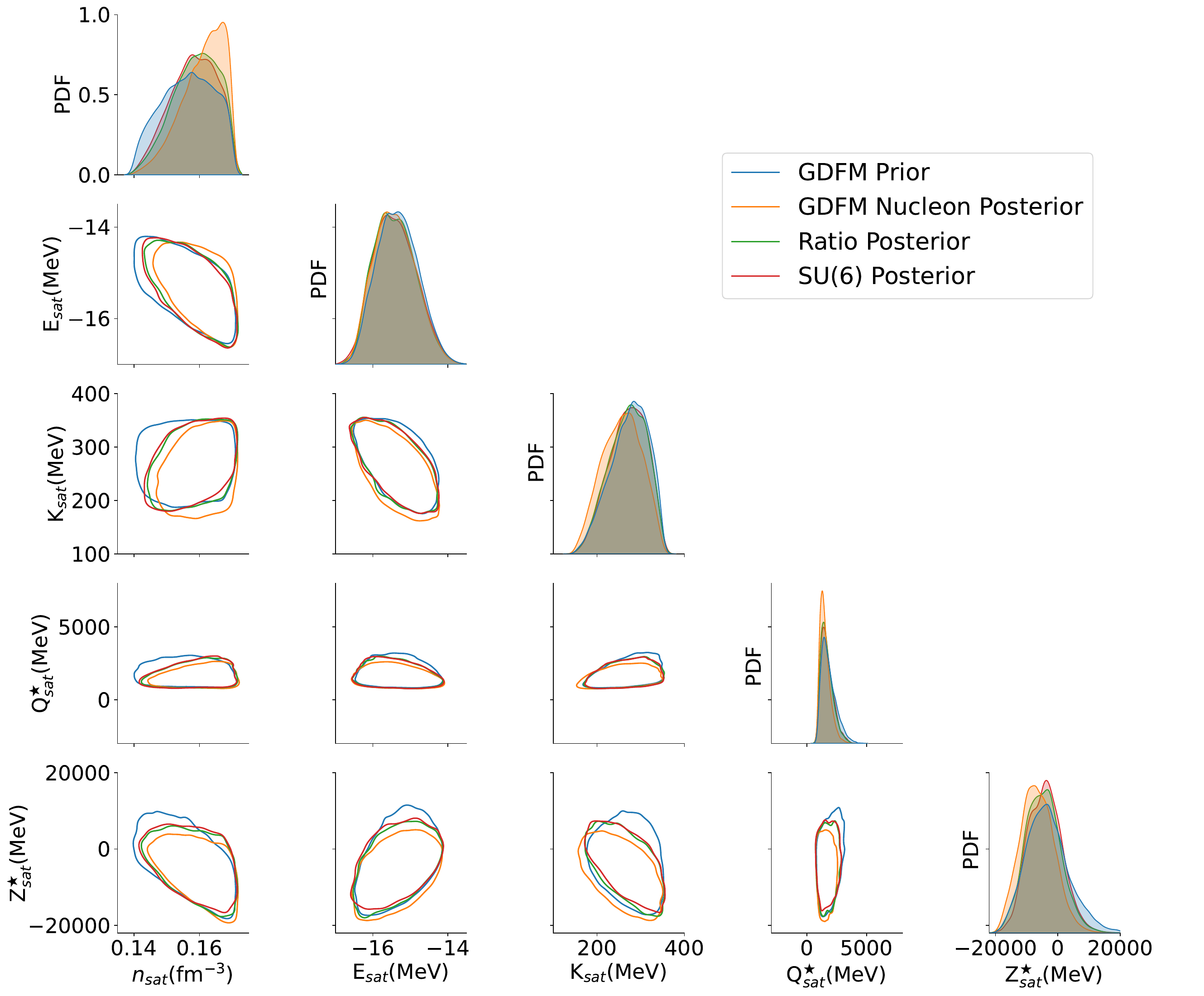}
\caption{{Probability distributions of} isoscalar NMPs for GDFM functional, along with the nucleonic posterior (as in Ref. \cite{Char:2025zdy}), SU(6) and Ratio hyperonic posteriors and their correlation contours within the 90\% CI. }
    \label{fig:nmp_sat}
\end{figure*}

\begin{figure*}
    \centering
\includegraphics[width=0.9\textwidth]{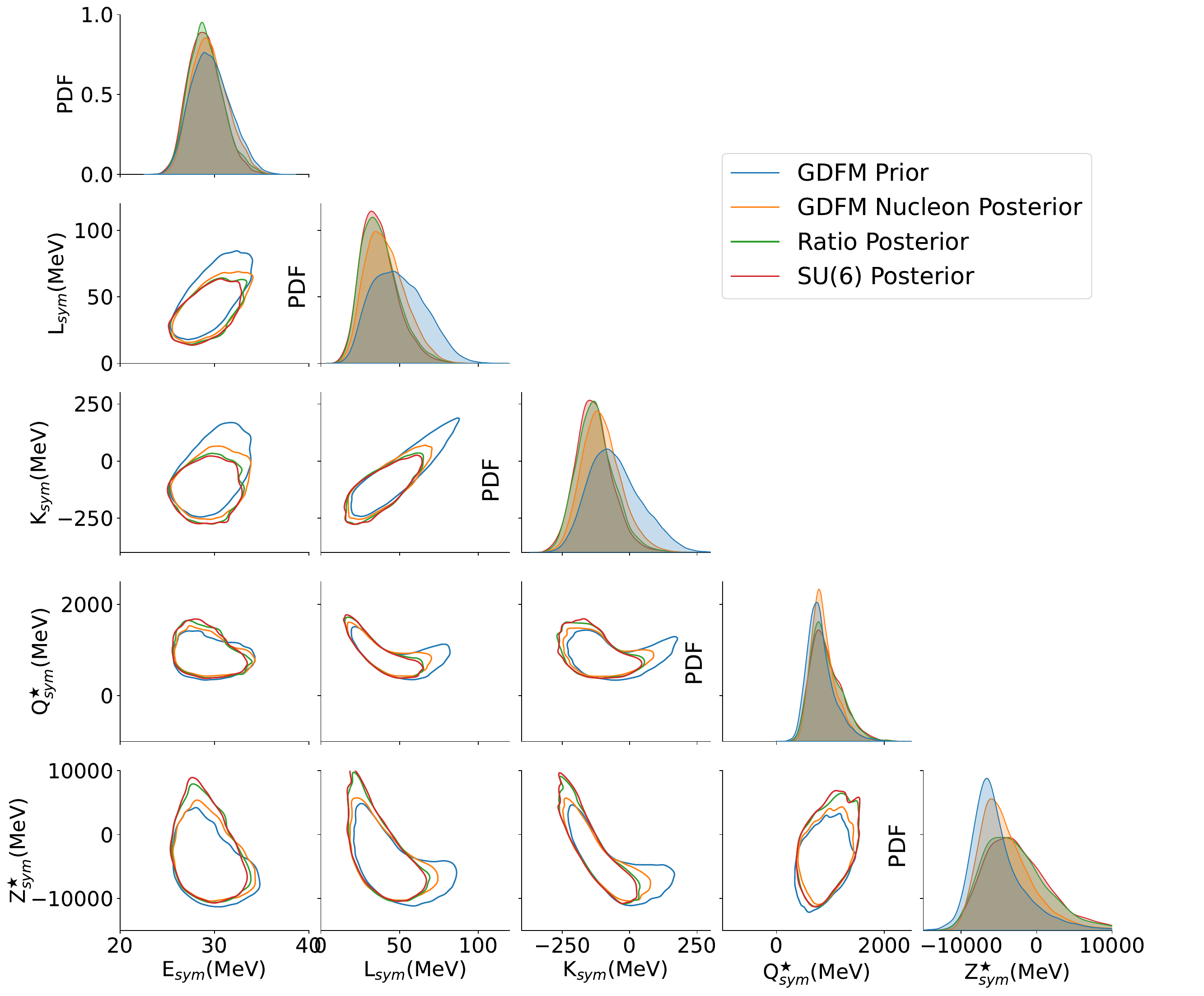}
\caption{The same as Fig. \ref{fig:nmp_sat} but for the isovector NMPs.}
    \label{fig:nmp_sym}
\end{figure*}

In figures \ref{fig:nmp_sat} and \ref{fig:nmp_sym}, we have shown the isoscalar and isovector parameters for the priors and posteriors of GDFM along with SU(6) and Ratio hyperonic posteriors.

The isocalar and isovector parameters are defined at saturation for symmetric matter. At the density, hyperons do not appear. Hence there is no direct impact on their calculations when we consider hyperons. However, the effect may come through the imposition of astrophysical constraints of pulsar mass measurements. Since the softening of the EOS due to the hyperons reduces the maximum mass, the EOSs with smaller values of incompressibility, $K_{\mathrm{sat}}$, are not preferred. We can see from Figure \ref{fig:nmp_sat} that the isoscalar contours for the Ratio and SU(6) are slightly shifted to the opposite direction of the nucleons. For the isovector parameters, we see in Figure \ref{fig:nmp_sym} slight reduction of the upper boundaries for $E_{\mathrm{sym}}$, $L_{\mathrm{sym}}$, and $K_{\mathrm{sym}}$ distributions. The ranges of $Q^*_{\mathrm{sym}}$ remains almost the same while the $Z^*_{\mathrm{sym}}$ distribution expands toward the higher values.

\bibliography{biblio}
\end{document}